\documentclass[12pt]{article}
\newcommand\mysection{\setcounter{equation}{0}\section}
\def\pw#1{^{#1}}
\def\rat#1#2{\mbox{\small $\frac{#1}{#2}$}}
\def\MSbar{\relax\ifmmode{\overline{\rm MS}}\else{$\overline{\rm MS}${ }}\fi}
\renewcommand{\theequation}{\thesection.\arabic{equation}}
\newcounter{hran}
\renewcommand{\thehran}{\thesection.\arabic{hran}}
\def\bmini{\setcounter{hran}{\value{equation}}
\refstepcounter{hran}\setcounter{equation}{0}
\renewcommand{\theequation}{\thehran\alph{equation}}\begin{eqnarray}}

\def\bminiG#1{\setcounter{hran}{\value{equation}}
\refstepcounter{hran}\setcounter{equation}{-1}
\renewcommand{\theequation}{\thehran\alph{equation}}
\refstepcounter{equation}\label{#1}\begin{eqnarray}}
\def\emini{\end{eqnarray}\relax\setcounter{equation}{\value{hran}}\renewcommand
{\theequation}{\thesection.\arabic{equation}}}

\def\la{\mathrel{\mathpalette\fun <}}
\def\ga{\mathrel{\mathpalette\fun >}}
\def\fun#1#2{\lower3.6pt\vbox{\baselineskip0pt\lineskip.9pt
  \ialign{$\mathsurround=0pt#1\hfil##\hfil$\crcr#2\crcr\sim\crcr}}}
\def\others{{\it et al.}}
\def\ie{\hbox{\it i.e.}{}}      \def\etc{\hbox{\it etc.}{ }}
\def\eg{\hbox{\it e.g.}{}}      \def\cf{\hbox{\it cf.}{ }}
\def\eV{{\rm e\kern-0.12em V}}            \def\MeV{{\rm M}\eV}
 \def\GeV{{\rm G}\eV} \def\TeV{{\rm T}\eV}
\def\half{{\textstyle {1\over2}}}

\def \alpi {{\alpha_s \over \pi}}
\def \al {\relax\ifmmode{\alpha}\else{$\alpha${ }}\fi}
\def \be {\relax\ifmmode{\beta}\else{$\beta${ }}\fi}

\def\abs#1{\left| #1\right|}
\def\lrang#1{\left\langle #1 \right\rangle}
\def\ben{\begin{enumerate}}  \def\een{\end{enumerate}}
\def\bit{\begin{itemize}}    \def\eit{\end{itemize}}
\def\beq{\begin{equation}}   \def\eeq{\end{equation}}
\def\beeq{\begin{eqnarray}}  \def\eeeq{\end{eqnarray}}
\def\bq{\begin{quote}}       \def\eq{\end{quote}}
\def\kp{\relax\ifmmode{k_\perp}\else{$k_\perp${ }}\fi}
\def\kps{\relax\ifmmode{k_\perp\pw2}\else{$k_\perp\pw2${ }}\fi}
\def \as{\relax\ifmmode\alpha_s\else{$\alpha_s${ }}\fi}
\def\alef{\relax\ifmmode{\ase}
 \else{$\ase${ }}\fi}
\def\aef{\relax\ifmmode {a_{\mbox{\tiny eff}}}
\else{$a_{\mbox{\tiny  eff}}${ }}\fi}
\def \pt{\relax\ifmmode{p_t}\else{$p_t${ }}\fi}

\def\ad{\relax\ifmmode\gamma\else{$\gamma${ }}\fi}
\def\eps{\epsilon}

\def\ee{\relax\ifmmode{e\pw+e\pw-}\else{${e\pw+e\pw-}${ }}\fi}
\def\qq{\relax\ifmmode{q\overline{q}}\else{$q\overline{q}${ }}\fi}
\def\QQ{\relax\ifmmode{Q\overline{Q}}\else{$Q\overline{Q}${ }}\fi}
\def\qqg{\relax\ifmmode{q\overline{q}g}\else{$q\overline{q}g${ }}\fi}

\def\br{brems\-strah\-lung\ }    

\newskip\humongous \humongous=0pt plus 1000pt minus 1000pt
\def\caja{\mathsurround=0pt}
\def\eqalign#1{\,\vcenter{\openup1\jot
\caja   \ialign{\strut \hfil$\displaystyle{##}$&$
\displaystyle{{}##}$\hfil\crcr#1\crcr}}\,}
\newif\ifdtup

\def\eqal2#1{\,\vcenter{\openup1\jot
\caja   \ialign{\strut \hfil$\displaystyle{##}$&\hfil$
\displaystyle{{}##}$\hfil &$
\displaystyle{{}##}$\hfil\crcr#1\crcr}}\,}

 \def\refup#1{~$^{\cite{#1}}$}
 \def\refupd#1#2{~$^{\citd{#1}{#2}}$}

 \def\cite#1{[\ref{#1}]}
 \def\citd#1#2{[\ref{#1},\ref{#2}]}
 
 \def\citm#1#2{[\ref{#1}--\ref{#2}]}
\def\ib#1#2#3{{\em ibid.}~\underline{#1} (19#3) #2}
\def\np#1#2#3{{\em Nucl.Phys.}~\underline{B#1} (19#3) #2}
\def\pl#1#2#3{{\em Phys.Lett.}~\underline{#1B} (19#3) #2}
\def\pr#1#2#3{{\em Phys.Rev.}~\underline{D#1} (19#3) #2}
\def\prep#1#2#3{{\em Phys.Rep.}~\underline{#1} (19#3) #2}
\def\prl#1#2#3{{\em Phys.Rev.Lett.}~\underline{#1} (19#3) #2}

\def\sjnp#1#2#3{{\em Sov.J.Nucl.Phys.}~\underline{#1} (19#3) #2}
\def\spj#1#2#3{{\em Sov.Phys.JETP}\/~\underline{#1} (19#3) #2}
\def\spu#1#2#3{{\em Sov.Phys.Usp.}\/~\underline{#1} (19#3) #2}
\def\jl#1#2#3{{\em JETP Lett.}~\underline{#1} (19#3) #2}
\def\zp#1#2#3{{\em Zeit.Phys.}~\underline{C#1} (19#3) #2}
\def\st{\sigma_{\mbox{\small tot}}}
\def\cF{{\cal{F}}}
\def\cK{{\cal{K}}}
\def\cV{{\cal{V}}}
\def\cR{{\cal{R}}}
\def\cQ{{\cal{Q}}}
\def\cO#1{{\cal{O}}\left(#1\right)}
\def\tt{$t\bar{t}$}

\newcommand{\naive}{na$\ddot{\imath}$ve}
\def\ase{\as^{\mbox{\tiny eff}}}
\def\beql#1{\beq\label{#1}}
\def\baeq{\begin{appeq}}     \def\eaeq{\end{appeq}}
\def\baeeq{\begin{appeeq}}   \def\eaeeq{\end{appeeq}}
\newenvironment{appeq}{\beq}{\eeq}
\newenvironment{appeeq}{\beeq}{\eeeq}
\def\bAPP#1#2{
 \markright{APPENDIX #1}
 \addcontentsline{toc}{section}{Appendix #1: #2}
 \medskip \medskip
 \begin{center}   {\bf\LARGE Appendix #1 :}{\quad\Large\bf #2} \end{center}
 \renewcommand{\thesection}{#1.\arabic{section}}
 \setcounter{section}{0} \setcounter{equation}{0}
 \renewcommand{\thehran}{#1.\arabic{hran}}
\renewenvironment{appeq}
  { \renewcommand{\theequation}{#1.\arabic{equation}} \beq }{\eeq}
\renewenvironment{appeeq}
  {  \renewcommand{\theequation}{#1.\arabic{equation}} \beeq }{\eeeq}
\nopagebreak \noindent}
\def\eAPP{\renewcommand{\thehran}{\thesection.\arabic{hran}}}

\begin{document}
\begin{titlepage}
\vfill

\LARGE
\begin{center}
{\bf Specific Features of Heavy Quark Production.
 LPHD approach to heavy particle spectra}

\vspace{.3in}
\large  Yu.L. Dokshitzer
\footnote{on leave of absence from the  Institute for Nuclear Physics,
St. Petersburg, Gatchina 188350, Russia}
\vspace{0.1in}

Theoretical Physics Division, CERN,\\
CH--1211, Geneva 23, Switzerland,\\

\vspace{0.2in}
 V.A. Khoze $^1$

Centre for Particle Theory, University of Durham,\\ Durham DH1 3LE, UK,

\vspace{0.2in}
S.I. Troyan

Institute for Nuclear Physics,\\  St. Petersburg, Gatchina 188350, Russia

\vfill

\begin{abstract}
\noindent
Perturbative QCD formula for inclusive energy spectra of heavy quarks
from heavy quark initiated jets
which takes into account collinear and/or soft logarithms in all orders,
the exact first order result and two-loop effects
is applied to distributions of heavy flavoured
hadrons in the framework of the LPHD concept.
\end{abstract}
\vfill
\end{center}
\normalsize
\end{titlepage}
\setcounter{footnote}{0}

\mysection{Introduction}
Heavy flavour physics is now extensively studied experimentally
at both \ee and hadronic colliders.
Experiments at $Z^0$ have led to the availability of new data on the
profiles of jets initiated by heavy quarks\refupd{frs}{exp}.
Further progress is expected from the measurements at LEP 2 and, especially,
at a future linear \ee collider.
The principal physics issues of these studies are related not only to
testing the fundamental aspects of QCD, but also to their large potential
importance
for measurements of heavy particle properties: lifetimes, spatial oscillations
of flavour, searching for CP-violating effects in their decays \etc
Properties of b--initiated jets are of primary importance for analysis of the
final state structure in   \tt\ production processes.
A detailed knowledge of the b-jet profile is also essential for the Higgs
search  strategy.

The physics of heavy quarks has always been considered as one of the best
testing grounds for QCD. Despite this, not many attempts to derive
self-consistent perturbative (PT) results for the profiles of heavy quark
jets have appeared so far.
This paper is aimed at a condensed presentation of the part of the results
of the long-term Leningrad/St.Petersburg QCD group
project\footnote{for some basic references see \citm{r3}{r8}}
that concerns PT analysis of the inclusive energy spectra of heavy quarks.

The selection of this subject is motivated by the following topical questions:
\bit
\item[i)]
To what extent can heavy quark distribution be treated as a purely
perturbative (Infrared Safe) quantity?
\item[ii)]
Can we describe the gross features of jets initiated by heavy quarks without
invoking phenomenological fragmentation schemes?
\item[iii)]
What kind of information can be extracted from the measurements of heavy quark
energy spectra and, in particular, from the scaling violation effects?
\eit
There are two ingredients of the standard Renormalization Group (RG) approach
to the description of the energy spectra of heavy flavoured hadrons $H_Q$.
Here one starts from a phenomenological fragmentation function for the
\beq\label{hadr}
      Q(x)\to H_Q(x_H)
\eeq transition and then traces its evolution with
the annihilation energy $W$ by means of PT QCD.
Realistic fragmentation functions\refup{fragm} exhibit a
parton-model-motivated maximum at (hereafter $M$ is the heavy quark mass)
$$
1-\frac{x_H}{x} \sim \frac{\mbox{const}}{M} \>.
$$
Taking gluon radiation at the PT-stage of evolution into account
would then induce scaling violations that soften
the hadron spectrum by broadening (and damping) the original maximum
and shifting its position to larger values of $1\!-\!x_H$ with $W$ increasing.

This approach has been successfully tried by B.~Mele and
P.~Nason\refup{MeNa}, who have studied the effects of multiple soft gluon
radiation and have been looking for realistic fragmentation functions to
describe the present day situation with heavy particle spectra, and to make
reliable predictions for the future.
Being formally well justified,
such an approach, however, basically disregards the effects that the finite
quark mass produces on the accompanying QCD \br pattern,
since the $W$-evolution by itself
is insensitive to $M$ (save the power suppressed $M^2/W^2$ corrections).

At the same time, as far as one may consider $M$ a sufficiently large
momentum scale, it is tempting to carry out a program of deriving the
predictions that would keep under PT control, as much as possible,
the dependence of $H_Q$
distributions on the quark mass.

In Section 2 we present and discuss the PT formula for inclusive
quark spectra, $D(x;W,M)$,
that in the QED case would
give an unambiguous all-order-improved
{\em absolute}\/ prediction for the inclusive energy distribution of muons
produced in
\beq\label{eemm}
\ee \to \mu^\pm(x)\> \mu^\mp + \ldots
\eeq
Given an explicit dependence on $m_\mu$, one may theoretically
compare, say,  the $\mu$ and $\tau$ spectra, a goal that, generally speaking,
can not be achieved within the standard ``evolutionary'' framework.

In Section 3 the topical questions listed above are addressed.
We discuss the problem of ``infrared instability'' of the PT quark spectra
and suggest a solution of this formal difficulty based upon the notion of
an ``infrared regular'' effective coupling.
Having regularized the PT expression in this
way we observe that the gluon \br effects
(Sudakov suppression of the quasi-elastic kinematics $x\!\to\!1$)
lead to particle spectra that have a similar shape
to the non-PT fragmentation function.

An emphasis of the r\^ole of PT dynamics has been successfully tested in
studies of light hadron distributions in QCD jets.
Prompted by the success of the MLLA-LPHD approach\refup{MLLA_LPHD},
we consider an option of describing the gross features of heavy quark
initiated jets entirely by means of PT QCD,
without invoking any fragmentation hypothesis.

The LPHD philosophy would encourage us to continue the PT description of an
inclusive quantity down to small momentum scales, and with the hope that
such an extrapolation of the quark-gluon language
would be dual to the sum over all possible hadronic excitations.
For the problem under consideration, such an approach could
describe the energy fraction distribution averaged over heavy-flavoured
hadron states, the mixture that naturally appears, \eg, in the study of
inclusive hard leptons.

In Section 4, we report results on the comparison of
the PT predictions with the measured $c$ and $b$ energy losses at different
annihilation energies.
Fitting the rate of scaling violation in $\lrang{x}(W)$ results in the
QCD scale parameter $\Lambda$ that, after being translated into
the popular \MSbar scheme, agrees with the values extracted from
other studies.
On the other hand, the {\em absolute}\/ values of $\lrang{x_{c,b}}$
at a given energy are most sensitive to the behaviour of the radiation
intensity, say, below 1--2\GeV.

Limited experimental information does not at present allow us
to disentangle various possible shapes of \alef near the origin.
At the same time, a quite substantial difference between $\lrang{x_{c}}$ and
$\lrang{x_{b}}$
fits nicely into PT-controlled mass dependence,
provided the characteristic integral of the effective coupling
over the ``confinement'' region assumes a fixed value,
$$
\int_0^{2\GeV}dk\>
\frac{\ase(k)}{\pi}\approx 0.38 \GeV\,
$$
(with $k$ the linear momentum variable).
Given this value, one may predict the differential $c$ and $b$
energy distributions at arbitrary $W$, predictions which prove to be
unaffected in practice by our ignorance of the detailed behaviour of
\alef in the small momentum region.

In Section~5 the results of the study are discussed and conclusions drawn.

Detailed derivation of the master formula for the inclusive
heavy quark energy spectrum is given in Appendices.
In Appendix~A the problem of running coupling in the perturbative
radiator is dealt with.
Appendix~B is devoted to the second loop effects in the anomalous dimension.

\mysection{\label{PTsec} Perturbative Energy Spectrum of Leading Quarks}
We denote the cms annihilation energy by $W$,
and $M \equiv m\cdot W$ is the quark mass.

\noindent
Here we describe perturbative QCD result for the
inclusive energy spectrum of heavy quarks, $D(x;W,M)$, $x\equiv 2E_Q/W$,
produced in  $\ee\to Q(x) + \bar{Q}+$ light partons.

This function has the following properties:
\bit
\item it embodies the exact first order result\refup{BK}
$D^{(1)}={\as}\cdot f(x;m)$
       and, as a consequence,
\item it has the correct threshold behaviour at $1-2m \ll 1$;
\item   in the relativistic limit $m\ll1$,
    it accounts for all significant logarithmically enhanced contributions
    in high orders, including
\ben
\item
 running coupling effects,
\item
 the two-loop anomalous dimension and
\item
 the proper coefficient function with exponentiated Sudakov-type logs,
which are essential in the quasi-elastic kinematics,
$(1\!-\!x)\ll1$;
\een
\item it takes into full account the controllable dependence on
the heavy quark mass that makes possible a comparison between the spectra
of $b$ and (directly produced) $c$ quarks.
\eit

\noindent
By ``threshold behaviour'' we mean here the kinematical region of
non-relativistic quarks, $\abs{W\!-\!2M}\sim M$, in which gluon \br acquires
additional dipole suppression.
At the same time we will not account for the Coulomb effects that
would essentially modify the production cross section near the actual
threshold, $W\!-\!2M\ll M$\refup{thresh}.

We are mainly interested in the ``leading'' energy region of $x$
of the order of (or close to) unity.
For this reason we disregard in what follows
potentially copious production of \QQ pairs via
secondary gluon splitting\footnote{apart from the integral effect
of the unregistered
pairs embodied into the running coupling, see below},
\ie\ {\em singlet}\/ sea contribution,
as well as the specific $Q\to Q+\QQ$ transition that appears
in the {\em non-singlet}\/ anomalous dimension beyond the first
loop (for a review see \cite{Guido} and references therein).

Therefore the quark distribution we define below will satisfy the sum rule
\beq\label{sumrule}
 \int_{2m}^1 dx\> D(x;W,M)\>=\> 1\>.
\eeq
The accuracy of the PT prediction described in this paper is
\beq\label{acc}
  D(x;W,M) \cdot\left[\, 1+\cO{\as^2}\, \right] \>.
\eeq
This expansion is {\em uniform}\/ in $x$, that is, the $\as^2$ correction
does not blow up  (neither as a power, nor logarithmically)
when $(1\!-\!x)\to 0$.

\subsection{Definition of the leading quark distribution.}
The inclusive cross section for single particle production in \ee annihilation
is conveniently written in terms of structure functions as,
\eg, \refup{Guido}

\beq\label{angint}
\sigma_0^{-1}  \frac{d\sigma}{dx  }
= 3 \cF^+_L(x) + \cF^+_2(x)
=  \cF^+_L(x) + \cF^+_T(x)\>,
\eeq
with  $\sigma_0$ the standard Born cross section factor.

We define the normalized inclusive energy spectrum as
\beq\eqalign{\label{Ddef}
D(x;W,M) &=  \st^{-1}\> \frac{d\sigma}{dx} \>\>
\equiv \>\> \int_\Gamma \frac{dj}{2\pi i}\>\> x^{-j}\> D_j(W,M) \>;
}\eeq
where the contour $\Gamma$ runs parallel to the imaginary axis in the complex
moment $j$ plane.

Aiming at an {\em exact}\/ account of the first order QCD effects, we include
in $\st$ the first \as correction to the annihilation cross section,
that in the case of relativistic quarks reads
$$
 \st = \sigma_0 \left[\, 1+ \rat{3}{2}\, C_F\, a \>
       +\> \cO{a^2}\, \right]; \quad a=a(W^2) \equiv \frac{\as(W)}{2\pi}\>,
\quad C_F=\frac{N_c^2-1}{2N_c}=\frac43.
$$
With this definition, the structure functions $\cF$ in the
moment representation become
\beq\label{stfun}\eqalign{
 \cF_\al(j) &= c_\al(j)\> \left(1-\rat32 C_Fa\, [\,j^{-1}-1\,]\right)
\> D_j \cdot \left[\, 1 + \cO{a^2}\, \right]\>, \cr
c_L(j) &= C_Fa\, j^{-1}\>, \quad
c_T(j) = 1 + \half C_Fa\,j^{-1} \>,\quad  c_2(j)= 1 - \rat32 C_Fa\,j^{-1} .
}\eeq

\subsection{The radiator.}
The integrand of (\ref{Ddef}) may be written in the exponential form as
\beq\label{Dj}
\ln {D}_j = \int_{2m}^1 dx\left[\,x^{j-1}-1\,\right]\frac{dw(x;W,M)}{dx} \>,
\eeq
where the ``radiator'' $dw/dx$ originates from the improved first order
gluon emission probability, in which finite mass and
running coupling effects have been
included\footnote{Detailed derivation of the radiator is presented in
Appendix~A; see also \cite{DKThq}}.
The corresponding expression reads
\bminiG{radiator}
\label{radad}
C_F^{-1}\> v\, \frac{dw}{dx} &=&  \int_{\kappa^2}^{\cQ^2} \frac{dt}{t}\,
\left\{   a(t) \left[\,\frac{2(x\!-\!2m^2)}{1-x}+\zeta^{-1}(1-x)\,\right]
- a'(t) (1-x) \>\>+\>\> a^2(t)\Delta^{(2)}(x)\, \right\} \qquad{} \\
\label{radcf}
 & + & \beta(x) \left\{ -\frac{2x}{1\!-\!x}\left[a(\cQ^2) + a(\kappa^2)\right]
 + \zeta^{-1} \, \frac{x\,(x\!-\!2m^2)}{2(1\!-\!x)}
 \left(\frac{1-x}{1\!-\!x\!+\!m^2}\right)^2 a(\cQ^2)  \,\right\}.
\emini
Here the two characteristic momentum scales have been introduced,
\beq\label{twoscales}\eqal2{
\cQ^2 \equiv& \cQ^2(x)
&= W^2 \cdot\frac{(1-x)^2}{1-x+m^2}\>z_0 \>, \cr
\kappa^2  \equiv& \kappa^2(x)
&= M^2\cdot  \frac{(1-x)^2}{z_0}  \>,
}\eeq
with
$$
 z_0 \equiv \half\left( x-2m^2+\sqrt{x^2-4m^2}\, \right)
\>=\>  x + \cO{m^2}\>.
$$
The following notations were also used:
\beq\eqalign{
m\equiv \frac{M}{W}\le\half\>; \quad v & \equiv \sqrt{1-4m^2} \>; \qquad
 \beta(x) \equiv  \sqrt{1-4m^2/x^2}\>,\quad 0\le\beta\le v\>;  \qquad
  \zeta=1+2m^2\>;\cr
 a'(k^2) & = \frac{d}{d\ln k^2}\, a(k^2)\>; \qquad
 a(k^2)\equiv \frac{\as(k)}{2\pi}\>.
}\eeq
The radiator (\ref{radiator}) vanishes at
$$
 x= x_{min}\equiv 2m\>; \quad \left(\> \beta(x_{min})=0\,, \>\>
\cQ^2= WM\frac{(1-2m)^2}{1-m} = \kappa^2 \>\right),
$$
thus justifying the lower kinematical limit in eq.(\ref{Dj}).

\paragraph{Relativistic Approximation.}
In the relativistic approximation, ${m\ll1}$, we set
$ v=\beta=\zeta=1 $
to get a simplified expression
\bminiG{radrel}
&& \eqalign{
C_F^{-1}\frac{dw}{dx} &=   \int^{\cQ^2}_{\kappa^2} \frac{dt}{t}\> \left\{
a(t)P(x)-a'(t)\,(1\!-\!x)+a^2(t)\Delta^{(2)}(x)\,\right\}
 \cr  &
+  a(\cQ^2) \left\{ \frac{-2x}{1\!-\!x} + \frac{x^2}{2(1\!-\!x)} \right\}
\>+\>  a(\kappa^2) \left\{ \frac{-2x}{1\!-\!x}\right\}\>;} \\
&& \cQ^2 = W^2x(1-x)\>, \quad \kappa^2=M^2(1-x)^2/x\>,
\emini
where
\beq\label{GLAP}
P(x)=\frac{1+x^2}{1-x}\>.
\eeq
The integration variable $t$  determines the physical hardness scale of the
running coupling, and is related to the transverse momentum of the
radiation\footnote{For the space-like evolution similar relation
holds\refup{DS} with $x^{-1}$ substituted for $x$ in (\ref{scale})}.
In the dominant integration region,
\beq\label{scale}
 \kps\ll W^2\>, \quad t=x\cdot\kps\>.
\eeq
The lower limit $t\ge\kappa^2$ sets the boundary for the essential gluon
emission angles,
$$
t\sim E_Q\frac{2}{W} \cdot (\omega_g\Theta)^2 \>\ge\> \kappa^2 \sim M^2
\frac{\omega_g^2}{E_Q} \>\frac{2}{W}
\quad \Longrightarrow \quad \Theta \ge M/E_Q \equiv \Theta_0\>.
$$
This restriction manifests the ``dead cone'' phenomenon
characteristic for \br off a massive particle.
It is largely responsible for the differences between radiative particle
production in jets produced by a light and a heavy quark
(excluding the decay products of the latter), see \citm{r4}{r7}.

\subsection{Logs and their exponentiation.}
Singularities of the radiator (\ref{radiator}) at $x\!=\!1$,
when driven through the inverse Mellin transform (\ref{Ddef}), (\ref{Dj}),
give rise to $\ln(1-x)$ terms.
Bearing this in mind, one may represent the term-by-term structure of the
radiator by the following symbolic expression,
\bminiG{symbexp}
\label{adsymb}
(\ref{radad})  &\Longrightarrow&
a(CS+S^2)\>+\> aC\>+\>  a^2C\>+\>  \left[\,a^2C+ \underline{a^2(CS+S^2)}\,
\right] \>; \\
\label{cfsymb}
(\ref{radcf})  &\Longrightarrow&  aS\>+\>aS\>+\>aS\>,
\emini
where $C=\ln W/M$ and $S=\ln1/(1\!-\!x)$
represent large logs that usually reflect enhancements due to quasi-collinear
and soft gluon radiation respectively.
As we shall see shortly, under the ``physical'' definition of the QCD coupling
the last 2nd order term in (\ref{adsymb}) is free from the double-log
contribution (the underlined piece) so that  the ``convergence''
of the PT expansion (\ref{symbexp}) is improved.

Exponentiation of collinear logs
follows from the general factorization theorem\footnote{this applies
to $C$-contaminated terms of (\ref{adsymb}) as well as to the last contribution
in (\ref{cfsymb}), that actually is due to {\em hard}\/ gluons
{\em collinear}\/ to the $\bar{Q}$ momentum
(the ``backward jet'' correction\refup{DKThq})}.
Arguments in favour of exponentiation of $aS^2 + aS$ contributions
according to (\ref{Dj}) have been given in \cite{DKThq},
motivated by the factorization property of soft \br.
Exponentiation of non-logarithmic $\cO{a}$ corrections,
corresponding to the terms in the non-integral part of the radiator
(\ref{radcf}) that are {\em regular}\/  at $(1\!-\!x)\to0$,
is questionable (indeed wrong) and unnecessary within the
accuracy adopted (see (\ref{acc})).
Keeping such terms in the exponent is a matter of choice.
We do so just to simplify the result and make
the normalization sum rule (\ref{sumrule}) automatically satisfied,
$D_{j=1}\equiv 1$.

When translated into the standard language of the RG motivated approach,
the integral part of the radiator (\ref{radrel}) may be said to embody
the anomalous dimension together with (a part of) the correction to the
hard cross section due to the $x$-dependent factor in the upper
integration limit:
\beq\label{badexp}
 \int^{W^2x(1-x)} \frac{dt}{t}  \, a(t)P(x)
\approx  \int^{W^2} \frac{dt}{t} \, a(t)P(x)
\>+\> a(W^2) P(x) \ln [x(1\!-\!x)]  \>+\> \cO{\frac{a^2\ln^2(1-x)}{1-x}}\>.
\eeq
Non-uniformity of such an expansion with respect to the potentially large
logarithm, $\ln(1-x)$,
explains why we preserve the exact kinematical limits in
(\ref{radiator}).
By neglecting the last term in (\ref{badexp}), one would lose
control of the $a^2 S^3$ terms.
Such contributions formally belong to the 2nd order correction
to the hard cross section,
and therefore lie beyond the reach of the two-loop RG analysis.
In the meantime, such an ignorance would undermine the possibility of
keeping track of essential first subleading corrections, $a^{n}\log^{2n-1}$,
at the level of running coupling effects in the quark form factor
(Sudakov suppression).

Let us stress that eqs.(\ref{Dj}), (\ref{radiator})
solve the problem of all-order resummation of soft radiation effects
both in the {\em hard cross section}\footnote{for a review of such
resummation programs see, \eg, \cite{Luca} and references therein}
and in the {\em coefficient function},
that is at low momentum scales $\kappa^2\propto M^2$.
In particular, the term in (\ref{radcf}) proportional to $a(\kappa^2)$
(which is $W$-independent and therefore gets lost
in the standard evolution approach)
accounts for the {\em exponentiated}\/ contribution from soft gluons at small
emission angles $\Theta<\Theta_0$, $\kps\ll M^2$
--- the dead cone subtraction effect.
Notice that the similar term proportional to $a(\cQ^2)$
(which does belong to the hard cross section corrections)
is due to the dead cone subtraction in the backward jet.

\subsection{Second loop effects.}
The preliminary result of \cite{DKThq} is improved by
taking into full account effects of the two-loop anomalous dimension
embodied in the $a'$ and $\Delta^{(2)}$ of (\ref{radad}).
We prefer to treat these two terms separately,
as the former naturally emerges in the dispersion relation approach
to definition of the running coupling\refup{GL}.
It shows that beyond the first loop
the ``soft'' and the ``hard'' parts of the GLAP splitting function (\ref{GLAP})
in the anomalous dimension
actually acquire different physical interaction strength\refup{DS},
$$
\ad^{(1)}= a\cdot\frac{1+x^2}{1-x} \>\>\Longrightarrow \>\> \ad^{(2)} =
a\cdot\frac{2x}{1-x} \>+\> [a-a']\cdot(1-x)\>.
$$
The actual expression for $\Delta$ depends on the renormalization scheme
chosen for the one-loop coupling.
It may be fixed, \eg, by matching with the known two-loop \MSbar result
for the non-singlet fragmentation function.
To this end one has to
consider the $W$-dependent part of the relativistic radiator (\ref{radrel}),
\beq\label{simp}\eqalign{
C_F^{-1}\frac{dw}{dx} &=   \int^{W^2x(1-x)} \frac{dt}{t}\, \left\{
a_{\MSbar}(t)P(x)-a'(t)\,(1\!-\!x)+a^2(t)\Delta_{\MSbar}^{(2)}(x)\,\right\} \cr
& +   a(W^2) \left\{ \frac{-2x}{1\!-\!x} + \frac{x^2}{2(1\!-\!x)} \right\}
\>+\> \mbox{const} \>+\> \cO{a^2(W^2)}\>,
}\eeq
and,  with account of (\ref{stfun}) relating $D$ to the structure functions,
compare the scaling violation rate according to (\ref{Dj}), (\ref{simp})
with that computed in \citm{CFP}{secloop}.

By doing so (see Appendix~B) one arrives at
$( C_A\!=\!N_c\!=\!3\>,\>\> T_R=\half )$
\bminiG{DtKdef} \label{Dtdef}
\Delta_{\MSbar}^{(2)}(x)  &=&
P(x)\cdot \cK\>+\> \tilde{\Delta}^{(2)}(x) \>; \\[2mm]
\label{cKdef}
 {\cK} &=&  \left[\, C_A\left( \frac{67}{18} -\frac{\pi^2}{6}\right)
 -\frac{10}{9} n_fT_R
 \,\right]  = \left\{
\eqalign{ 4.565  & \quad (n_f=3) \,, \cr
          4.010  & \quad (n_f=4) \,, \cr
          3.454  & \quad (n_f=5) \,;
} \right.
\\
\label{Deltatdef}
  \tilde{\Delta}^{(2)}(x) &=&  C_F\cV(x) + \cR(x) \, .
\emini
$\cV$ term is responsible for the difference between the time-like
and space-like anomalous dimensions.
Explicit expressions for $\cV(x)$ and $\cR(x)$ are given in Appendix~B.

\paragraph{Introducing the Physical Coupling.}
In the $x\!\to\!1$ limit, $\Delta_{\MSbar}^{(2)}$ (\ref{DtKdef})
peaks together with $P(x)\propto (1\!-\!x)^{-1}$.
At the same time, $\tilde{\Delta}^{(2)}$ is less singular in this limit,
$$ \eqalign{
\left. \cV(x)  \right/{\Delta}^{(2)}(x)  & \propto (1-x) \ln(1-x) \>, \cr
\left. \cR(x)  \right/{\Delta}^{(2)}(x)  & \propto (1-x)^2 \>.
}$$
As far as large quark energies are concerned, $\ln1/(1\!-\!x)> 1$,
the $\tilde{\Delta}^{(2)}$ term constitutes a small correction.
Thus, the major part of the two-loop effect according to (\ref{DtKdef})
reduces to a finite renormalization of the leading term $aP(x)$.
This naturally suggests absorbing the correction
by introducing the ``physical'' effective coupling  according to
\bminiG{phcointr}
\label{twosch}
 a_{\MSbar}P(x) + a^2\Delta_{\MSbar}^{(2)}(x)
& =& \aef P(x) + a^2 \tilde{\Delta}^{(2)}(x)   \>+\>\cO{a^3}\>; \\
\label{phco}
\aef  = a_{\MSbar}\left( 1+ a\,\cK \right)\>, \quad &&
\quad a_{\MSbar} \approx \aef \left( 1- \aef \,\cK \right) .
\emini
Eq.(\ref{phco}) relates the scale parameters  $\Lambda$ of the two schemes.
This relation within two-loop accuracy reads
\beql{lambeff}\eqalign{
\ln \Lambda_{\mbox{\tiny eff}} &
  = \ln \Lambda_{\MSbar} + \frac{\cK}{\beta}  \>\>=\>\>
\ln \Lambda_{\MSbar} \>+\>
\frac{\left( \frac{67}{18} -\frac{\pi^2}{6}\right) C_A
 -\frac{10}9 n_fT_R }{\frac{11}{3}C_A - \frac43 n_fT_R}\>; \cr
\Lambda_{\mbox{\tiny eff}} &= \Lambda_{\MSbar} \times \left\{
\eqal2{ &1.66 & \quad (n_f=3)\>, \cr
        &1.57  & \quad (n_f=5)\>.}      \right.}
\eeq
Our $\Lambda_{\mbox{\tiny eff}}$ coincides with the
$\Lambda_{\mbox{\small MC}}$ introduced by
Catani,  Marchesini and Webber in \cite{CMW90}.

\paragraph{Quark Thresholds in the Running Coupling.}
One additional comment is in order concerning the essence of \alef.
In spite of the formal equivalence of the two representations (\ref{twosch}),
the former one is physically preferable since
$ \tilde{\Delta}^{(2)}(x) $
is free from an ill-defined quantity $n_f$ representing
the number of ``active'' quark flavours.
The $t$-integration in (\ref{radiator}) runs over a broad region
that is sliced by the finite quark mass scales.
Naturally, one has to increment $n_f$ when passing such a scale,
\eg\ $n_f\!=\!3\to n_f\!=\!4$  when going through  $t\!\approx\! 1.5\GeV$.
$\Lambda_{\MSbar}$ has to be redefined by adjusting to a new
$\beta$-function value, and $\Delta_{\MSbar}$ should be changed accordingly.
The problem appears to be entirely technical, however, and
reflects a deficiency of the \MSbar prescription as one of the schemes
based upon the dimensional regularization technique that
eventually treats fermions as massless particles:
The discontinuous behaviour of $\Lambda$ and $\Delta$ gets compensated
in the physical combination (\ref{twosch}) that emerges in
the full two-loop anomalous dimension.
It is worthwhile to notice that our convention
($n_f$ independent $ \tilde{\Delta}^{(2)}(x) $, the scale parameter $\Lambda$
defined by (\ref{lambeff}))
is in accord with the BLM prescription\refup{BLM} for optimizing the choice
of ``physical coupling''.

For the argument sake, one may once again invoke the QED example (\ref{eemm}).
In this context, an application of the \MSbar scheme would reveal the same
problem with the lepton thresholds ($e,\mu,\tau$ \etc), while \aef
would be nothing but the {\em physical}\/ running QED coupling, the one
given by the Euclidean photon renormalization function $Z_3(t)$
that is unambiguously linked by the dispersion relation
to fermion pair production (with the finite fermion mass effects fully
included).

The improved one-loop expression for \aef accounting for finite quark mass
effects reads (for a detailed discussion see \cite{DS})
\bminiG{Tcoup}
\aef^{-1}(Q^2) &=&    \beta_\ell \ln\frac{Q}{\mu}
- \frac{2T_R}{3}\sum_i \Pi\left(1+\frac{4M_i^2}{Q^2}\right)
  \>+\>   \aef^{-1}(\mu^2) \>,
\qquad (\mu\ll M_i)\>;\\
\label{Pias}
 \Pi
&\approx& \left\{  \eqal2{ & \ln  ({Q^2}/{M_i^2}) - \rat53 \>,
\quad & Q\gg M_i \>;\cr   &  {2Q^2}/{5M_i^2}\>, \quad & Q\ll M_i \>,
} \right.
\emini
where $\beta_\ell$ accounts for gluon and massless quark contributions
($\beta_\ell=b_{[n_f\!=\!3]}=9$)  and $\Pi$ is the
standard fermion loop polarization operator known from the mid-fifties
(see below eq.(\ref{Tec})).
According to (\ref{Pias}), heavy flavours {\em decouple}\/ at low momentum
scales (as they ought to).
In the ``ultraviolet'' regime, $\Pi$
acquires a constant subtraction from the leading log behaviour, which
term, if treated as the two-loop correction, explains the $n_f$ dependent
piece of the $\cK$ factor (\ref{cKdef}) that enters the relation
(\ref{phco}) between the physical and the \MSbar couplings:
$-\rat{10}{9}T_R = \rat{4}{3}T_R\cdot (-\rat{5}{6})$.

\paragraph{$\tilde{\Delta}^{(2)}$ is Practically Negligible.}
As an estimate of the magnitude of the $\tilde{\Delta}^{(2)}$ correction
we present its contribution to the second moment $j\!=\!2$
that describes quark energy losses:
\beq\label{dsdef} \eqalign{
& C_F\left\{ a_{\MSbar} P_j\, (
1\,+\, a \,\cK ) \>+\> a^2\tilde{\Delta}_j^{(2)} \right\}
\>=\> C_F \,P_j \left\{\, \aef \>+\>
a^2 \,\delta_j \, \right\} \>+\>\cO{a^3}; \cr
& \delta_2  =  \left[ \frac{21}{36}
+  \left(\frac{85}{36} -\frac{\pi^2}{3} \right)\right] C_F
+ \frac{31}{36}\left(C_A -2C_F\right)  =    - 0.1735\> .
}\eeq
Being numerically very small already for $j\!=\!2$, the relative correction
$\delta_j$ continues to fall as $j^{-1}$ with $j$ increasing.
Therefore in practice one may use (\ref{radiator}) with $\Delta^{(2)} \equiv0$,
provided the physical effective coupling (\ref{phco}) is used.

\subsection{Axial vs. Vector currents in \QQ production.}
The master equation (\ref{radiator}) has been written for the case of
\QQ production via the vector current, in which case
$$
\zeta= \zeta_V \equiv \frac{\sigma_{[V]}(v)}{\sigma_{[V]}(1)} =
 1+2m^2 = \half(3-v^2) \>.
$$
This factor tends to 1 in the relativistic limit $m\to 0$ but
depends otherwise on the production channel (see Appendix~A for details).
It is worthwhile to note that beyond the leading twist approximation,
generally speaking, there is no way to define the universal
{\em fragmentation}\/ distributions that would be independent of the
{\em production}\/ stage.
The origin of such non-universality (that shows up at the level of
$\as\, m^2\ln m^2$ terms) may be traced back to the process dependent
radiation of {\em hard} gluons,\footnote{
both the ``soft'' and ``semi-soft'' terms of the radiation spectrum,
$\propto\>\omega_g^{-1}\cdot d\omega_g$ and $\propto\>1\cdot d\omega_g$,
remain universal\refupd{LBK}{DKS}.}
$$
 {d\sigma}\>\propto\> \omega_g \cdot {d\omega_g} \>\propto\> (1-x)\,dx.
$$
In practice we have to deal with a mixture of vector and axial
production currents.
In the pure axial case $\zeta$ should be substituted by
$$
 \zeta_A \equiv \frac{\sigma_{[A]}(v)}{\sigma_{[A]}(1)} =
 1-4m^2 =  v^2 \>.
$$
In addition, the $\zeta^{-1}(1-x)$ term in (\ref{radiator})
acquires an extra mass correction factor $(1\!+\!2m^2)$.
Both this effect and the difference between $\zeta_V$ and $\zeta_A$ are
proportional to $m^2=(M/W)^2$.
We conclude that
\bit
\item
the difference between the radiation spectra in
$V$- and $A$-channels vanishes as $(1\!-\!v^2)$ in the relativistic case, while
\item
for non-relativistic quarks, $v\ll1$, the axial contribution to the cross
section is relatively suppressed as $A/V\sim v^2$.
\eit
Therefore (\ref{Ddef})--(\ref{radiator})
may be used in practice for the realistic $V\!+\!A$ mixture,
in particular, at the $Z^0$ peak.

\mysection{Hadronization and Infrared Finite \as }
In the QCD context the non-PT phenomena inevitably enter the game.
It is important to stress however that the strong interaction not only
determines the hadronization transition (\ref{hadr})
but also affects to some extent the quark evolution stage:
The high order effects embodied into the  {\em running coupling}\/
seem to undermine the very possibility of the PT analysis.

It is often believed that the large quark mass, $M\gg\Lambda$, provides
a natural cut-off, which keeps the relevant space-time region compact
enough to avoid the truly strong, non-PT interaction in a course of the
quark evolution.

To illustrate the point we invoke a rough estimate that follows immediately
from (\ref{Dj}) and is valid for numerically large $x$ values
in the double-log approximation\refup{DKThq},
\bminiG{Dapp}
D(x;W,M) \propto (1-x)^{C_F\Delta\xi-1}\>,
\eeeq
with $\Delta\xi$ the characteristic evolution integral
\beeq
\label{MGI64}
\Delta\xi = \int^{(1 - x)W^2}_{(1 - x)^2 M^2}\frac{d\kps}{\kps}
\> \frac{\as(\kp)}{\pi}\>.
\emini

In the fixed coupling approximation we would get
\beq
 D(x) \propto (1-x)^{\frac{C_F\as}{\pi}\ln\frac{W^2}{M^2}-1}
\cdot \exp\left\{ -\frac{C_F\as}{2\pi} \ln^2(1-x)\right\}.
\eeq
In the small coupling regime,
$C_F\frac{\as}{\pi}\ln\frac{W^2}{M^2}<1$, this distribution exhibits
a sharp peak at large $x$ followed by a steep fall off in the $x\to1$ limit
due to the double logarithmic Sudakov suppression.
Qualitatively similar shape for the heavy hadron spectrum due to
$Q\to H_Q$ transition one expects from the parton model
considerations\refup{leading} which have lead to the so called
Peterson fragmentation function\refup{fragm}
\beq\label{pet}
C_Q(x) \>=\>\frac{N}{x} \left[\> \frac{1-x}{x}
+ \frac{\eps_Q}{1\!-\!x}\>\right]^{-2} \>,\qquad
N=\frac{4{\eps_Q^{1/2}}}{\pi}\left[\, 1 + \cO{\eps_Q^{1/2}}\,\right].
\eeq
This function peaks at $1\!-\!x\approx{\eps^{1/2}}\ll 1$ leading to
\beq\label{petenlos}
\lrang{1-x}_{fragm} \>=\> {\eps_Q^{1/2}} \cdot
\frac{2}{\pi} \left(\ln\eps_Q^{-1}-1\right)
\left[\, 1+ \cO{\eps_Q^{1/2}}\,\right] \>\propto\>\eps_Q^{1/2} \>.
\eeq

\subsection{Peterson fragmentation function vs. integrated coupling.}
One may single out effects of the non-PT momentum region by factorizing the
$D$ spectrum in the moment representation (\ref{Dj})
into the product of the ``safe'' PT part and the ``confinement'' factor,
\bminiG{Dprodrad}
\label{Dprod}
D_j = D_j\left[\, \kps>\mu^2\,\right]  \>\times\> D_j^{(C)}\>.
\eeeq
In other words, we split the radiator into two pieces corresponding to large
and small transverse momentum regions ($t\approx\kps$ for $x$ close to 1),
\beeq
\label{Drad}
\frac{dw}{dx} &=& \frac{dw}{dx}\left[\, \kps>\mu^2\,\right] \>+\>
\left\{\frac{dw}{dx}\right\}^{(C)}\left[\, \kps\le\mu^2\,\right].
\emini
This formal separation becomes informative if one is allowed to choose the
boundary value $\mu$ well {\em below}\/ the quark mass scale
(\eg, $\mu\!=\!1\GeV$ providing $\mu/M\!\ll\!1$ for the $b$ quark case).
Within such a choice only $x$ close to 1 would contribute to $w^{(C)}$.
{} For illustrative purposes let us neglect subleading $\mu/M$
effects and retain the most singular term only to get an estimate
\bminiG{confradiator}
\left\{\frac{dw}{dx}\right\}^{(C)} \approx
\vartheta\left(\frac{\mu}{M}-(1\!-\!x)\right)\> \int_{[(1-x)M]^2}^{\mu^2}
\frac{dt}{t}\, a(t) \> \frac{2C_F}{1-x} \>+\>\ldots
\eeeq
One arrives at
\beeq\label{conffact}
\eqalign{
 \ln D_j^{(C)} &\approx
\int_0^{\mu/M}  {dz} \left[\,(1-z)^{j-1}-1\,\right]
\int_{[zM]^2}^{\mu^2} \frac{dt}{t}\>  a(t)\,\frac{2C_F}{z} \cr
&= 2C_F  \int_{0}^{\mu} \frac{dk}{k}\, \frac{\as(k)}{\pi}
\int_0^{k} \frac{du}{u}
\left[\, \left(1-\frac{u}{M}\right)^{j-1} -1\,\right] .
}\emini
It is important to stress that it is the first ``confinement insensitive''
factor of (\ref{Dprod}) only that depends on $W$.
Therefore,  the ratio of the moments
\bminiG{twopreds}
\label{pred1}
 \left. D_j(W,M) \right/ D_j(W_0,M)
\eeeq
as a function of $W$ and the heavy quark mass is expected to be an ``infrared
stable'' PT prediction.
Another message one receives observing the structure of the radiator
(\ref{radiator}) is that the ratio of the moments for
{\em different}\/ quarks,
\beeq
\label{pred2}
 \left. D_j(W,M_1) \right/ D_j(W,M_2)\,,
\emini
should tend to a $W$-independent confinement sensitive ({\em sic}\/!)
constant in the relativistic limit
$W\gg M_1, M_2$.

We proceed with the estimate of the ``confinement'' factor (\ref{conffact}).
As far as $\mu/M$ may be treated as a small parameter, for finite moments
$j\!\sim\!1$, $j\mu/M\ll1$,
\beq\label{finitej}
 \ln D_j^{(C)} =
-2C_F(j-1)\,\int_0^{\mu} \frac{dk}{M}\, \frac{\as(k)}{\pi}  \>
\left[\, 1 +\cO{\frac{jk}{M}}  \,\right]\approx
 -2C_F\,(j-1) \, \frac{\mu_\al}{M}\>,
\eeq
with
\beq\label{mualdef}
\mu_\al \equiv \int_0^{\mu} dk\> \frac{\as(k)}{\pi}\,.
\eeq
In particular, for the energy losses, $j\!=\!2$, one has
\beq\label{cnfenlos}
 \ln D_2^{(C)} \equiv \ln \lrang{x}^{(C)} =
 -2C_F\frac{\mu_\al}{M} \>\left[\,1+\cO{\frac{\mu}{M}} \,\right],\quad
\lrang{x}^{(C)} =  \exp\left\{-2C_F\frac{\mu_\al}{M}\right\}.
\eeq
If (\ref{finitej}) were applicable for {\em all}\/ $j$, the inverse Mellin
transform would result in a singular distribution
\beq
 D(x)^{(C)} = \delta\left(x - \lrang{x}^{C} \right).
\eeq
This singularity gets smeared when a proper treatment is given to the
large $j$ region.
To this end one may use a simple expression
that interpolates between (\ref{finitej}) and the correct logarithmic
asymptote of the ``confinement radiator'' in (\ref{confradiator}),
\beq\label{interpol}
 \ln D_j^{(C)} \approx -2C_F \int_0^\mu \frac{dk}{k}\, \frac{\as(k)}{\pi}
 \>\ln \left[\, 1+ \frac{k}{M}(j-1)\,\right] .
\eeq
As a result, a distribution emerges that peaks around $x=\lrang{x}^{(C)}$
and is rather similar in shape to (\ref{pet}).
The extreme $x\!\to\!1$ asymptote is determined by the region of
parametrically large moments $\lrang{j}\propto  (1\!-\!x)^{-1} \gg M/\mu$.
It is different for two regimes:
\bminiG{twoass}
\label{gz}
\as(0) >0\,, \quad
\ln D_j^{(C)}  \sim -C_F\frac{\as(0)}{\pi}\>\ln^2 \left( j\frac{\mu}{M} \right)
& \Longrightarrow &  D(x)\propto  \frac{\exp\left\{ - C_F \frac{\as(0)}{2\pi}
\ln^2(1\!-\!x)  \right\}}{ 1-x } ;\qquad{} \\
\label{ez}
\as(0) =0\,, \quad
\ln D_j^{(C)}  \sim -2 C_F \Xi_0 \> \ln \left( j\frac{\mu}{M} \right)\quad
& \Longrightarrow&   D(x)\propto (1-x)^{2C_F \Xi_0-1}\>,
\emini
where, in the latter case,
\beq\label{Anot}
 \Xi_0 \equiv  \int_0^\mu \frac{dk}{k}\>\frac{\as(k)}{\pi} \><\> \infty\,.
\eeq
Reproducing the concrete behaviour of the Peterson function (\ref{pet})
in the large $x$ limit,
$C(x)\sim (1\!-\!x)^2$,
would require $\Xi_0=\rat{9}{8}$ in the second regime (\ref{ez}).
We conclude that the particle distribution originating from
(\ref{confradiator}) is capable of reproducing
the gross features of the popular Peterson fragmentation function
(provided, naturally, that the notion of the infrared regular effective
coupling is implanted in the PT-motivated ``confinement'' radiator).

An important message comes from comparing the energy losses that occur
at the hadronization stage.
Based on the pick-up hadronization picture, the characteristic parameter
$\eps$ in (\ref{pet}) has been predicted to scale\refup{fragm} as
\beq\label{Mscaling}
 \eps_Q \approx \left(\frac{m_q}{M}\right)^2 \propto M^{-2},
\eeq
(with $m_q$ the quantity of the order of constituent light quark mass.)
Confronting (\ref{petenlos}) that (up to a logarithmic factor) scales as
$\sqrt{\eps_Q}$ with the PT prediction (\ref{cnfenlos}) we get
$$
 \lrang{1-x}_{fragm} \sim \sqrt{\eps_Q} \>\> \Longleftrightarrow\>\>
 1- \lrang{x}^{(C)} = 1- \exp\left\{-2C_F\frac{\mu_\al}{M}\right\}
\approx 2C_F\frac{\mu_\al}{M}\>,
$$
which justifies the expected scaling law\footnote{Similar behaviour was
advocated
recently by R.L. Jaffe and L. Randall\refup{Jaffe}
who have exploited  the difference between the
hadron and the heavy quark masses as a small expansion parameter.}.

It is worthwhile to remember that neither the Peterson function nor our
PT-motivated ``confinement'' distribution is an unambiguously defined object.
The former as an ``input'' for the evolution is by itself contaminated by
gluon radiation effects at the hard scale $t\!\sim\! M^2$
that are present even at moderate $W\ga 2M$.
On the other hand, $D[\kps\le\mu^2]$ crucially depends on an arbitrarily
introduced separation scale $\mu$ that disappears only in the product of
the factors responsible for ``PT'' and ``non-PT'' stages (\ref{Dprod}).
Nevertheless, bearing this in mind, one may still speak of
a direct correspondence between these two quantities, namely,
$C(x)$ in the $j$ representation and $D_j^{(C)}$ as given by
(\ref{confradiator}).

This means that instead of convoluting phenomenological $C(x)$ with the
$W$-dependent ``safe''   evolutionary  quark distribution one may
try to use consistently the pure PT description that would place no artificial
separator between the two stages of the hadroproduction.
{}From the first sight, one gains not much profit
substituting one  non-PT object --- the phenomenological fragmentation
function $C(x)$ --- by another unknown, namely, the behaviour of
the effective long-distance interaction strength $\as(k)$
(at, say,  $k\la 2\GeV$).
There is, however, an important physical difference between the two approaches:
$\as$ should be looked upon as an {\em universal}\/ process independent
quantity. Therefore quite substantial differences between inclusive spectra
of c- and b- flavoured hadrons should be under complete control according
to the explicit quark mass dependence embodied in the PT formulae.

\subsection{Modeling the coupling.}
To study {\em infrared sensitivity}\/ of PT results one can try
different shapes of the effective coupling or, equivalently,
different ways to extrapolate the characteristic function
\beql{xidef}
\xi(Q^2) = \int^{Q^2} \frac{dk^2}{k^2}\> \frac{\as(k)}{\pi}
\>+\> \mbox{const}
\eeq
to the ``confinement'' region of small $Q^2$.
Using the one-loop expression for the coupling,
\beq\label{aldef}
\frac{\as^{(1)\,}(k)}{\pi} \equiv 2a^{(1)\,}(k^2)\>=\>
\frac{4}{b\ln (k^2/\Lambda^2)}\>;\qquad
 b=\frac{11}3N_c-\frac23 n_f\>,
\eeq
for $\xi$ one gets
\beql{xiPT}
 \xi^{(1)\,}(k^2) = \frac4b \ln \ln\frac{k^2}{\Lambda^2} \>+\>\mbox{const}\,,
\eeq
which expression is defined only for $k>\Lambda$.

For the two-loop effective coupling we use the standard approximate expression
\bmini\label{twoloopec}
\as^{(2)}(k) = \as^{(1)}(k) \left( 1
- \frac{b_1 \ln\ln({k^2}/{\Lambda^2})}{4\pi\> b}\, \as^{(1)\,}(k) \right)
\eeeq
with
\beeq
 b_1 = \frac{34}{3} N_c^2 - \left(\frac{10}{3} N_c + 2C_F\right) n_f
\emini
and $\as^{(1)}$ given by the one-loop formula (\ref{aldef}).
Analytic expression for $\xi$ then reads
\beq\label{xitwo}
  \xi^{(2)\,}(k^2) = \frac4{b} \left( \ln L +\frac{b_1}{b^2} \frac{\ln L+1}{L}
\right)\>+\>\mbox{const}\,; \qquad L\equiv \ln \frac{k^2}{\Lambda^2}\>.
\eeq

\subsubsection{$F$-model. \label{ALPIsec}}
The simplest prescription which we refer below as the $F$-model consists of
{\em freezing}\/ the running coupling near the origin.
One follows the basic PT dependence
given by either (\ref{aldef}) or (\ref{twoloopec})
down to a certain point $k_c^2$ where the coupling reaches a given value
\bminiG{alpixi}
\label{cpar}
\frac{\as(k_c)}{\pi} = A \>,
\eeeq
and then keeps this value down to $k^2\!=\!0$.
$\xi$ then takes the form
\beeq
\eqal2{
          =&  \xi(k^2) - \xi(k_c^2)\>, \qquad  & k^2>k_c^2\>; \cr
 \xi(k^2) =&  \>A \,\ln \left(\left.{k^2}\right/\!{k_c^2}\right)\>,\qquad
& k^2<k_c^2\>,
}\emini
with $k_c$ related to $A$ by (\ref{cpar}).

\subsubsection{$G$-model.}
The set of $G_p$-models (Generalized shift models) gives another more flexible
example for the trial effective coupling.
It emerges when one regularizes the evolution function (\ref{xiPT}) as follows,
\bminiG{ksigen}
 \xi^{(1)\,}(k^2) = \frac4b\ln\ln\left(\frac{k^{2p}}{\Lambda^{2p}}
 + C_p\right) \>+\>\mbox{const}\>,\quad C_p\ge 1 \>,
\eeeq
which corresponds to the effective coupling
\beeq\label{asp}
\frac{\as^{(1)\,}(k)}{\pi} \equiv \frac{d\,\xi^{(1)\,}(k^2)}{d\ln k^2}
= \left[\>\frac{k^{2p}}{k^{2p}+C_p\,\Lambda^{2p}}\>\right] \cdot
\frac{4}{b} \>\frac{p}{\ln\left(k^{2p}/\Lambda^{2p} + C_p\right)}\>
\>.
\emini
This expression preserves the perturbative asymptotic form (\ref{aldef})
up to power corrections $\Lambda^{2p}/Q^{2p}$.
Notice that the effective coupling (\ref{asp}) with $C_p=1$ has a finite limit
$\as(0)/\pi=4p/b$, while for $C_p\!>\!1$ it vanishes in the origin.

\begin{figure}[htb]
\vspace{11.5cm}
\includegraphics{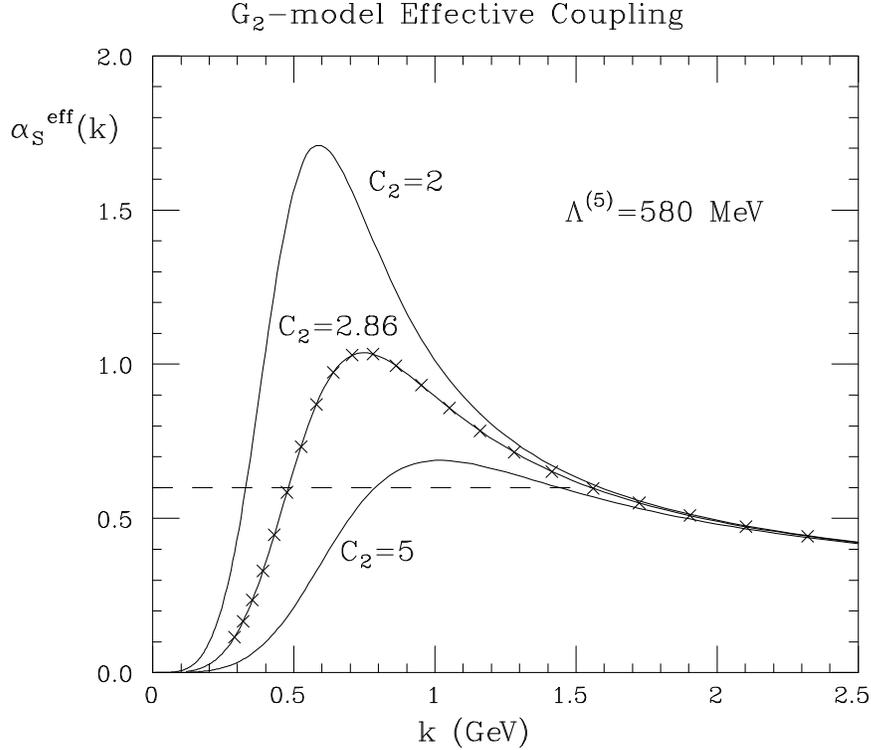}
\caption{Trial shapes of {\protect $\alpha_s^{\mbox{\tiny eff}}$}
in the {\protect$G_2$} model (two-loop, {\protect$n_f\!=\!5$}).
Marked by crosses is the best-fit coupling.  For comparison the best-fit
$F$-model coupling ($A$=0.19) is shown with dashed line.}\label{hqfig1}
\end{figure}

For the two-loop coupling one substitutes in (\ref{xitwo})
\bminiG{ksigentwo}
L \>\> \Longrightarrow \>\>
L_p= \frac1p \ln \left(\frac{Q^{2p}}{\Lambda^{2p}} + C_p \right),
\eeeq
which results in
\beeq\label{asptwo}
\frac{\as^{(2)\,}(k)}{\pi} \equiv \frac{d\,\xi^{(2)\,}(k^2)}{d\ln k^2}
= \left[\>\frac{k^{2p}}{k^{2p}+C_p\,\Lambda^{2p}}\>\right]
\cdot  \frac{4p}{b\,L_p} \left(1- \frac{b_1}{b^2}\frac{\ln L_p}{L_p} \right)\>.
\emini
Trial shapes of the effective coupling in the $G_2$ model
($G$-model with $p\!=\!2$ and the two-loop $\as$ with $n_f\!=\!5$
massless flavours)
are displayed in Fig.\ref{hqfig1} for different values of the parameter $C_2$.
Crosses mark the curve that provides the best fit to mean energy losses
(see below).

\subsubsection{Quark Thresholds.}
When it comes to an accurate account of heavy quark thresholds in $\as$
we modify the logarithmic denominator of $\as^{(1)}(k^2)$
in (\ref{twoloopec}) according to (\ref{Tcoup}) as follows
\bminiG{Tec}
 b_{[n_f\!=\!5]}\,\ln\frac{k^2}{\Lambda^2} \>\Longrightarrow\>
 b_{[n_f\!=\!3]}\,\ln\frac{k^2}{\Lambda^2} - \frac23 \left[\,
\Pi\left(1+\frac{4M_c^2}{k^2}\right) + \Pi\left(1+\frac{4M_b^2}{k^2}\right)
 \,\right]
\eeeq
with
\beeq
 \Pi(v^2) = \frac{v(3-v^2)}{2} \ln\frac{1+v}{1-v} +v^2-\frac83 \>.
\emini

Given infrared regular behaviour of $\ase$,
numerical evaluation of the inverse Mellin transform (\ref{Ddef})
with the PT radiator (\ref{radiator}) becomes straightforward.
Inclusive energy heavy quark spectrum obtained along these lines
is concentrated (has a sharp maximum)
near $x_Q=1$ at $W\ge 2M$ and softens due to gluon \br effects
with $W$ increasing.
Similar softening one achieves increasing radiation intensity in the PT
domain (by taking a larger value of $\Lambda$)
and/or at small momentum scales (by varying ``confinement parameter''
of the model; larger $A$, smaller $C_p$).

Since available experimental information on differential heavy quark spectra is
rather scarce at present (and possibly contradictory), we restrict ourselves
by considering the {\em mean energy losses}\/ which,
as we have discussed above,
can be studied to quantify the influence of non-PT effects.

The value of $\lrang{x_Q}\equiv D_2(W,M)$
shows up quite a strong variation with $A/C$.
However, from the general factorization argument
(\ref{pred1}) one would expect the {\em energy dependence}\/
of $\lrang{x_Q}$ to stay well under PT control.
As demonstrated in \cite{preprint}, the normalized quantity
\beql{normenlos}
  \left. \lrang{x_Q}(W) \right/ \lrang{x_Q}(W_0)
\eeq
is indeed practically insensitive to the variation
of the ``confinement parameter'' of the model.
Therefore the ratio (\ref{normenlos}) may be looked upon as an
{\em infrared stable}\/ quantity suitable for measuring $\Lambda$
as has been suggested by P.M\"attig\refup{frs}.

Position of the peak in the energy distribution seems to give another less
trivial
example of a stable prediction.
Once again, as in the case of $\lrang{x_Q}$, the {\em absolute}\/ value of
the peak position depends strongly on the chosen $A$ value.
At the same time the {\em normalized}\/ quantity
$\left. x_{\mbox{\tiny peak}}(W)\right/ x_{\mbox{\tiny peak}}(W_0)$
exhibits much weaker $A/C$-dependence than the PT-controlled dependence
on $\Lambda$\refupd{DKThq}{r8}.

Thus the $W$-evolution (scaling violation) in quark
energy losses $\lrang{x_Q}(W)$ allows one to extract the scale parameter
$\Lambda$ which value proves to be practically insensitive
to the adopted scheme of $\ase$ extrapolation.
At the same time the {\em absolute}\/ values of $\lrang{x_Q}$
are quite sensitive to the gross effective radiation intensity below
1--2$\GeV$ which makes it possible to quantify corresponding
``confinement parameter'' of the scheme and, thus, the shape of the coupling.

Worthwhile to notice that there is a natural theoretical scale the
``measurement'' of $\ase$ below 1--2$\GeV$ to be compared to.
As shown by Gribov\refup{BH}, in the presence of light quarks
colour confinement occurs when the effective coupling
(parameter $A$ of the $F$-model)
exceeds rather {\em small}\/ critical value
\beq\label{alcrit}
A \> >\>  \left\{\alpi\right\}^{crit}
= C_F^{-1}\left[\>1-\sqrt{\textstyle \frac23} \>\right] \approx 0.14\>.
\eeq
Thus, within the Gribov confinement scenario an interesting possibility arises.
Namely, if phenomenological $\ase$ extracted from the data
does exceed $\as^{crit}$ but remains numerically small this would provide
a better understanding of the PT approach to multiple hadroproduction
in hard processes.

\mysection{Numerical Analysis of Energy Losses}
In this Section we
compare experimental data with the generalized PT prediction which embodies
the notion of the infrared regular effective QCD coupling.
As an input we take the world average values\refup{frs}
of $\lrang{x_Q}$ listed in Table~\ref{tab1}.

\begin{table}[htb]
\begin{center}
\begin{tabular}{l c | c | c}
& process & $W_{cm}(\GeV)$ & $\lrang{x_Q}$ \\ \hline
1. &      & 10.4 & $0.727 \pm 0.014                       
$ \\[1mm]
2. &$c\to D^*+\ldots$   & 30   & $0.587 \pm 0.015         
$ \\
                3. &      & 91   & $0.508  \pm 0.009        
$ \\ \hline\hline
   4. &$c\to\ell+\ldots$ & 57.8 & $0.541\pm0.036 $
 \\
   5. &$c\to\ell+\ldots$ & 91 & $0.522 \pm 0.022         
$ \\ \hline
6. &$b\to\ell+\ldots$ & 29--35 & $0.789 \pm 0.022        
$ \\
   7. &$b\to\ell+\ldots$ & 91 & $0.699\pm 0.009          
$  \\ \hline
\end{tabular}
\end{center}
\caption{Experimental measurements of $\lrang{x_Q}(W,M)$ }
\label{tab1}
\end{table}

\noindent
Errors have been evaluated by taking
statistical and systematic errors in quadrature.
First 3 entries stand for the direct production of $D^*$ mesons at
different centre-of-mass energies; the last 4 data for the mean {\em quark}\/
energy have been extracted by unfolding the inclusive lepton ($e,\mu$)
spectra from heavy $Q$ decays.

As mentioned above, the PT approach advocated in this paper
can not pretend to fully describe {\em exclusive}\/ heavy hadron spectra
(with $D^*$ an example).
Our treatment of the hadronization stage that implicitly appeals to
duality arguments makes it plausible to rather apply this approach
to inclusive quantities such as the lepton energy distributions.
Nonetheless, for lack of anything better, we take the measured
mean energy of $D^*$ as a representative of $\lrang{x_Q}$ to be
compared directly with the PT motivated prediction for
the quark energy losses.

The preliminary analysis has shown\refupd{DKThq}{r8}
that an independent fitting of the $D$ and $L$(epton) data results
in the best-fit curves $A_D(\Lambda)$ and $A_L(\Lambda)$ that
{\em cross}\/ just at the best-fit value of $\Lambda$.
This was the observation that motivated us to look upon $\ase$ as a
process independent quantity to confront the $c$ and $b$ measurements
1--7 with a unique one-parameter PT prediction\footnote{in spite of the fact
that such a \naive\ approach suggests the same theoretical expectation for
the two physically different data \# 3 and \# 5. }.

Hereafter we fix heavy quark masses to be
\beq\label{quarkmasses}
  M_c = 1.5\> \GeV\>, \qquad M_b = 4.75\> \GeV\,.
\eeq
(Sensitivity to the $b$--quark mass will be discussed below.)

\subsection{Fitting mean quark energies.}
\begin{figure}[htb]
\vspace{11.5cm}
\includegraphics{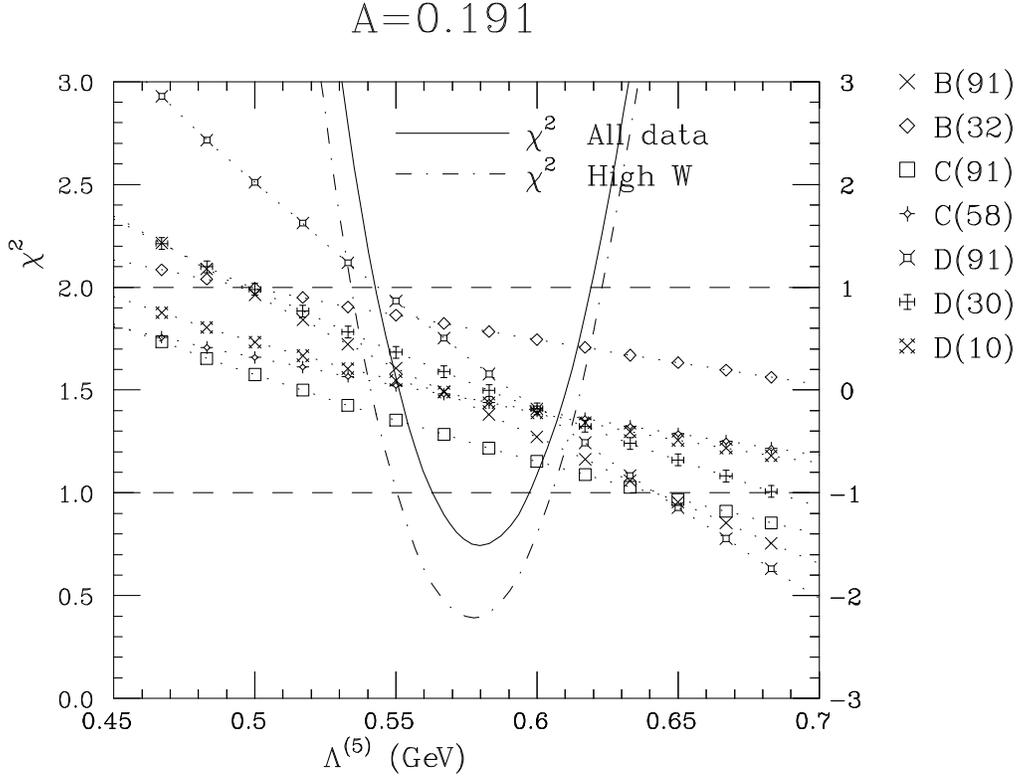}
\caption{$\Lambda$--dependence of the $F$--model fit to mean energy losses
(two-loop $\ase$ with $n_f\!=\!5$).
The right scale shows normalized deviation between a theoretical
prediction and an experimental datum (\protect\ref{whatplot}).
Dashed lines mark the $1\sigma$ band.
Solid and dash-dotted lines show the values of $\chi^2$
(against the left scale)
for all data and the high-$W$ data sample correspondingly. }\label{hqfig2}
\end{figure}

{Fig.\protect\ref{hqfig2}}
shows the quality of the fit to 7 separate
data of the Table~\ref{tab1} together with the total $\chi^2$
as a function of $\Lambda^{(5)}$ in the $G_2$-model with $C_2=2.86$
(the best-fit value).
Some explanation is in order.
In this Figure (and similar plots below) for each
datum the ratio is displayed
\beql{whatplot}
   \frac{\mbox{theor.} - \mbox{exp.}}{\mbox{exp.error}}
\eeq
against the right vertical scale.
Dashed horizontal lines mark $1\sigma$ levels for
each single datum.
The two curves (which should be read out against the left scale)
show the squared deviation of the points (\ref{whatplot}) from the
median, that is, total $\chi^2$.
The solid curve sums up all 7 data, while the
dash-dotted one accumulates the ``high $W$'' data only.
The ``high $W$'' sample we define excluding the entries \#1 and \#6
which correspond to the quark mass-to-energy ratios
$$
  {M}/{E} \>\approx\>  {M_c}/{5\GeV} \sim  {M_b}/{15\GeV}
 \>\approx\>   1/3 \>.
$$
These two entries are subject to significant non-relativistic corrections.
The fact that the two fits are consistent
is good news:
it shows that the non-relativistic effects have been properly
taken into account\footnote{the relativistic version of (2.4--7)
reported earlier\refup{DKThq} failed to properly embody the
``$b$ at 32'' datum \protect\# 6}
in the PT radiator (\ref{radiator}).

\begin{figure}[htb]
\vspace{14.5cm}
\includegraphics{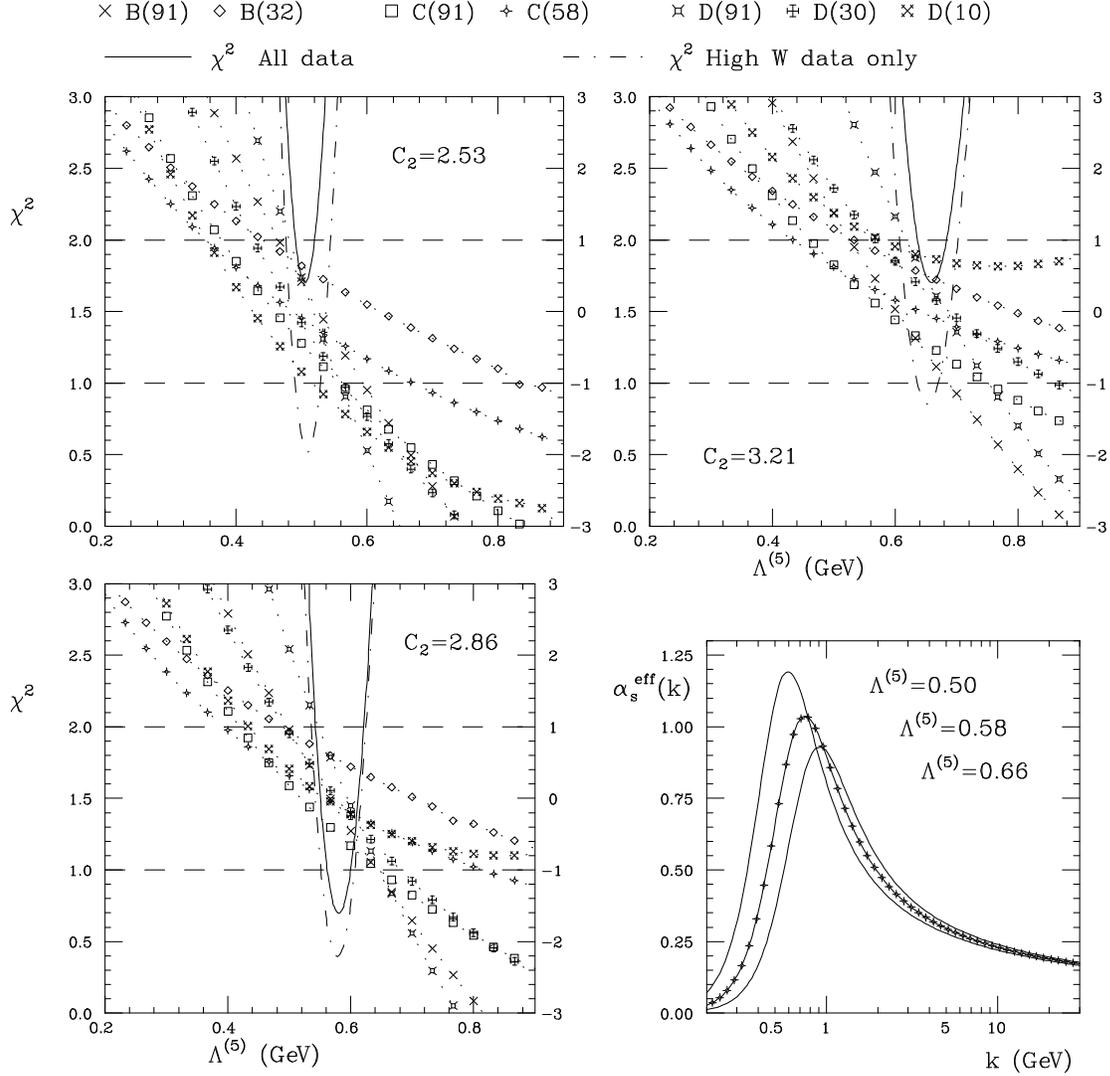}
\caption{
On sensitivity of the $G_2$-model fit to the shape of $\protect\ase$
in the origin  (two-loop, $\protect n_f\!=\!5$).
The values of $C_2$ in the upper plots correspond to one standard deviation
from the best-fit (bottom-left). Bottom-right graph displays the
margin in low momentum behaviour of $\ase$. }
\label{hqfig3}
\end{figure}

\paragraph{Fig.\protect\ref{hqfig3}}
demonstrates sensitivity of PT description to the ``confinement parameter''.
Here we have chosen the $G_2$--model for a change.
The first thing to be noticed
is that with the best-fit parameter $C_2=2.86$ one obtains
the same value $\Lambda^{(5)}\approx 580$~$\MeV$,
comparable quality of the fit $\chi^2_{\min}\approx 0.7$
and even the same dynamics of each of 7 data
as in the above $F$--model description (Fig.\ref{hqfig2}).
In the upper part of this plot two marginal values of $C_2$ are also shown
which correspond to one standard deviation from the total 7--fit:
$
\chi^2_{\min}(2.53)=\chi^2_{\min}(3.21)=\chi^2_{\min}(2.86)+1
$.

\begin{figure}[htb]
\begin{minipage}{3.0 in}
\vspace{9 cm}
\includegraphics{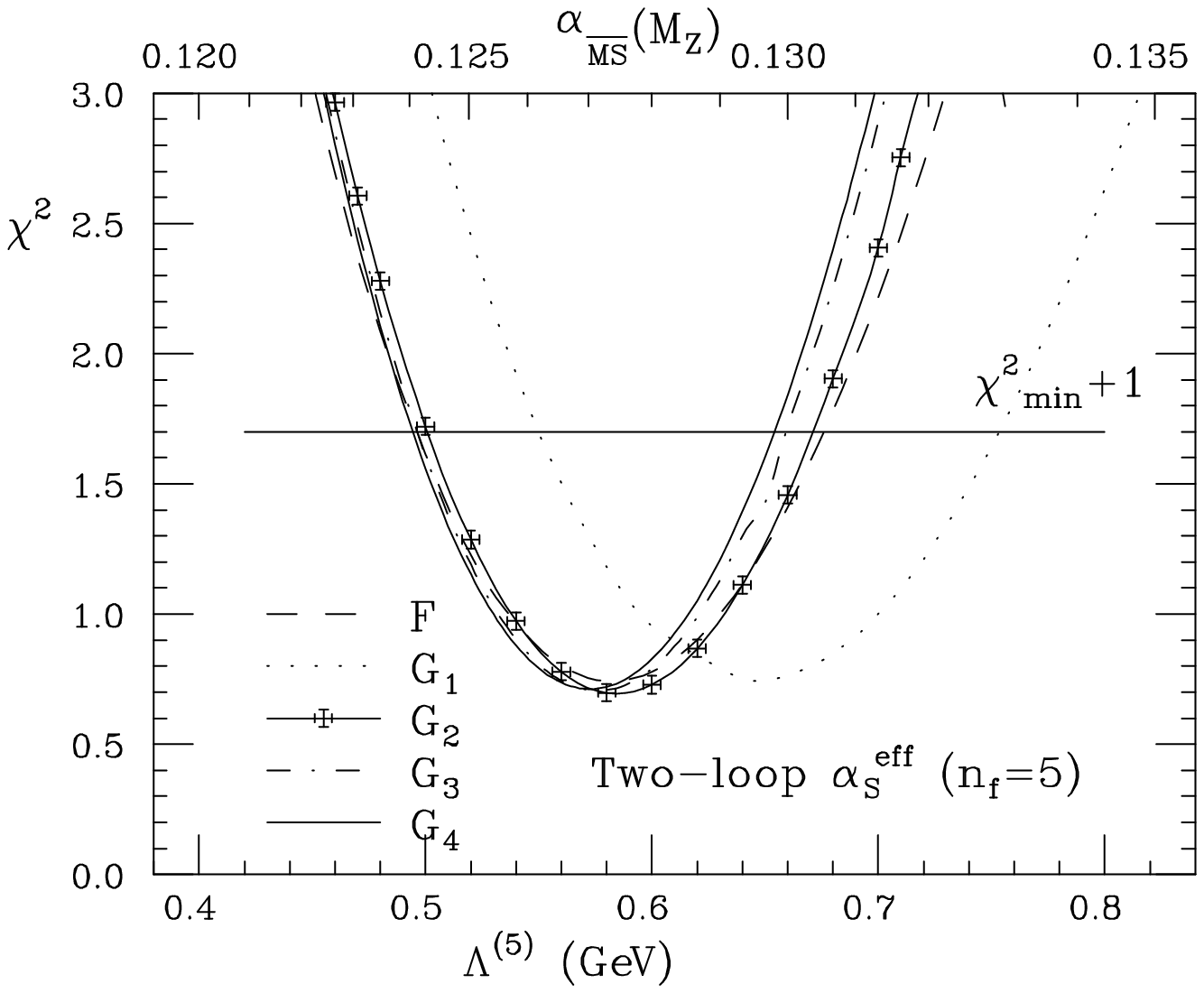}
\caption{One-parameter fits to energy losses within different models for
$\ase$.}
\label{hqfig4}
\end{minipage}
\quad
\begin{minipage}{3.4 in}
\vspace{9 cm}
\includegraphics{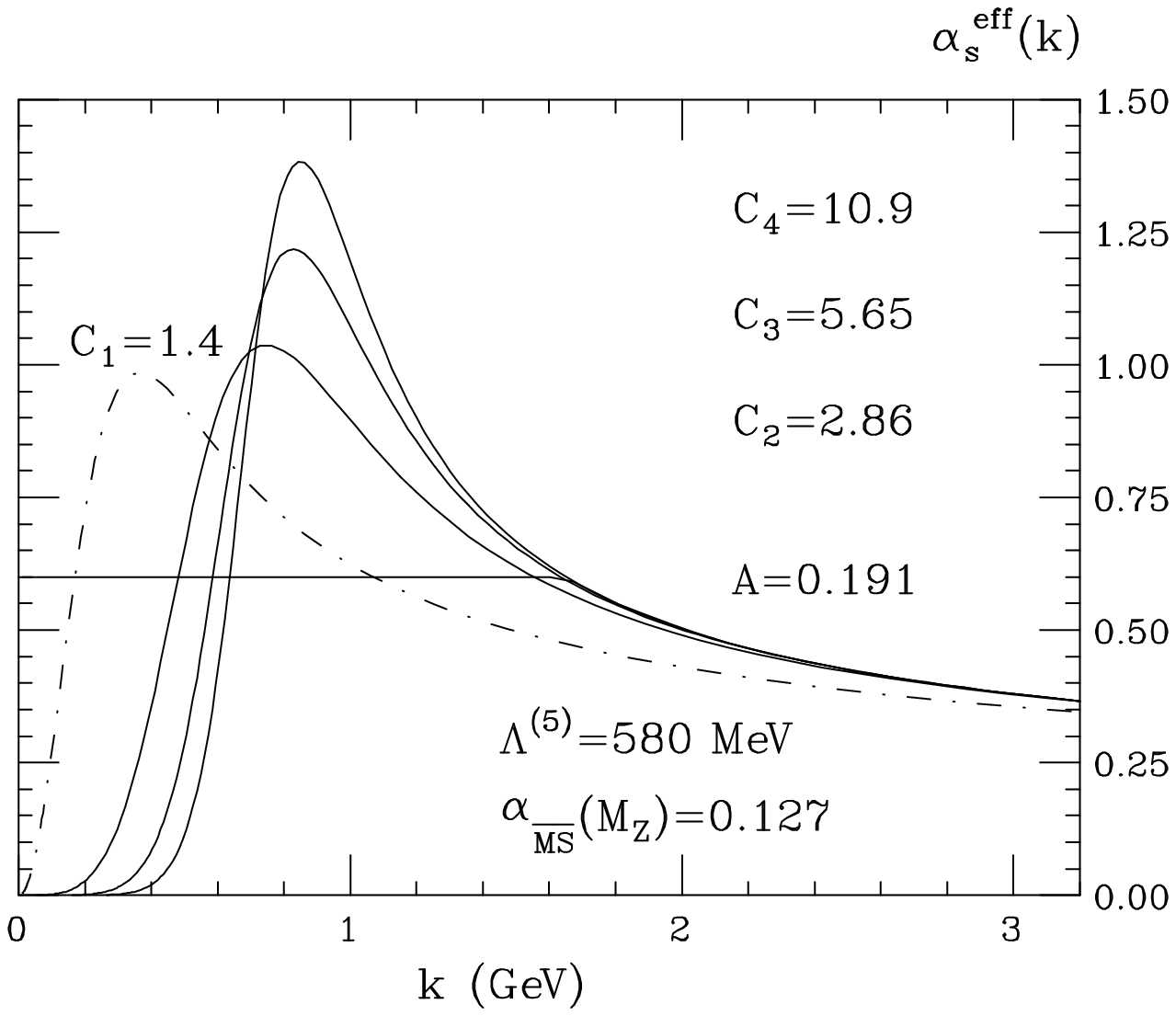}
\caption{Best-fit  $\ase$ for $\Lambda^{(5)}=580\MeV$.
$G_1$--model underestimates radiation above $1.5\GeV$.}
\label{hqfig5}
\end{minipage}
\end{figure}

\paragraph{In Fig.\protect\ref{hqfig4}}
comparison is made of the quality of the fits within different models
for the effective coupling (two-loop, $n_f\!\!=\!\!5$).
For each value of $\Lambda$ the $A/C$ parameter has been adjusted to minimize
the error (one-parameter fit).
The upper scale shows corresponding values of $\alpha_{\MSbar}$ at LEP
recalculated from $\ase$ with use of the relation (\ref{phco}).

It is worthwhile to notice some peculiarity of the $G_1$--model.
This model is ``too soft'' in a sense that it induces the {\em negative}\/
preasymptotic power term  $\propto k^{-2}$ in $\ase(k)$,
which correction suppresses $\ase$ in a relatively high momentum region.

\paragraph{Fig.\protect\ref{hqfig5}}
illustrates this peculiarity.
Here the couplings corresponding
to the best-fit $A/C$ values for $\Lambda^{(5)}=580$~\MeV are compared.
Solid lines ($F$, $G_{2-4}$) correspond to $\chi^2\approx 0.7$.
In the $G_1$--model shown by dash-dotted line ($\chi^2\approx1.2$)
$\ase$ stays noticeably smaller above $1.5\GeV$
before the perturbative logarithmic regime sets up and all the models merge.

As a result, to compensate for reduced radiation intensity
the best-fit $\Lambda$ value (and thus $\alpha(M_Z)$)
within the $G_1$--model
tends to be larger compared to ``sharp'' models $F$, $G_2,\ldots$

Leaving $G_1$ aside we conclude that both the quality of the fit
and the value of $\Lambda$ the ``sharp'' models point at,
hardly exhibit any model dependence.
{}From Fig.\ref{hqfig4} (see also Fig.\ref{hqfig3}) we deduce
\bminiG{loop2ress}
 \label{loop2resL}
 \Lambda^{(5)} &=& 580 \pm 80\> \MeV\>.
\eeeq
Being translated into the $\MSbar$ parameter this gives
\beeq
 \label{loop2resA}
 \alpha_{\MSbar}(M_Z) &=&  0.127 \pm 0.003\,.
\emini
The error here is purely statistical (one standard deviation).

In what follows we shall be using the 0.003 shift in $\alpha_{\MSbar}(M_Z)$
induced by the $G_1$--model (see Fig.\ref{hqfig4} and Table~2 below)
as a rough estimate of systematic uncertainty
due to possible ``soft'' preasymptotic power effects in the running
coupling.

\begin{figure}[htb]
\vspace{10.5cm}
\includegraphics{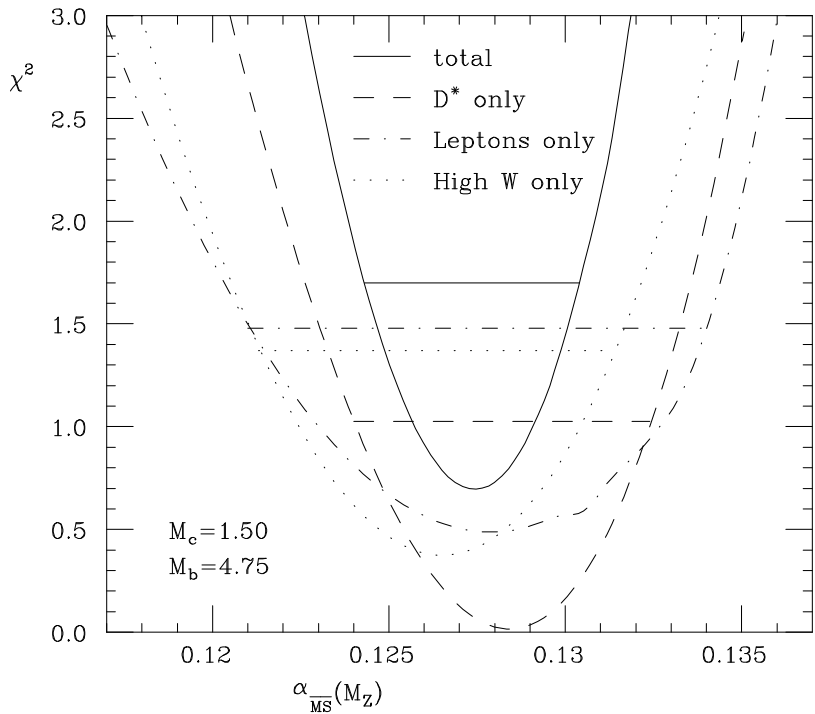}
\caption{On consistency between the fits to different subsets of data on mean
energy losses (two-loop, $n_f\!=\!5$).
Horizontal lines mark one standard deviation levels.
}\label{hqfig6}
\end{figure}

\paragraph{Fig.\protect\ref{hqfig6}}
demonstrates consistency of the total $G_2$--model fit with fits
to various subsets of data: $D^*$ (items 1--3 of Table~1),
``leptons'' (4--7), high $W$ (items 2--5, 7).
Here $\chi^2$ is plotted against the reference value $\alpha_{\MSbar}(M_Z)$.

$D^*$ data are more restrictive since they have smaller experimental errors
than inclusive lepton measurements.
Low--$W$ points $b\to$lepton+\ldots at 32 (\#6) and, especially,
$D^*$ at 10 (\#1)
are quite important as they provide lever arm for scaling violation.
Inclusion of these two measurements does not spoil the fit,
$\chi^2/5$ (total fit) $\approx$ $\chi^2/3$ (high--$W$),
but increases its quality reducing statistical error by factor 2.

\subsection{Energy ratios and $\Lambda$ determination.}
\begin{figure}[htb]
\vspace{11.5cm}
\includegraphics{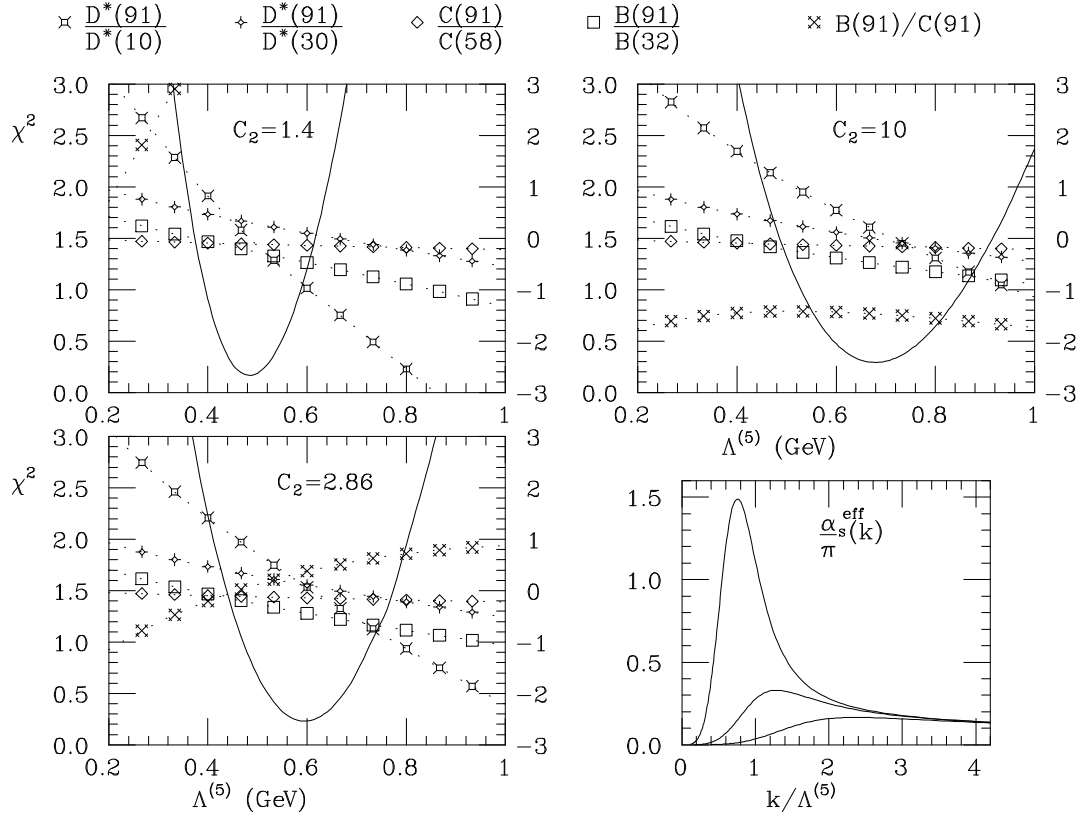}
\caption{
$C_2$-dependence of energy ratios ($G_2$-model, two-loop, $\protect n_f\!=\!5$)
with $\chi^2$ for the first four ratios shown by
solid lines and
correponding shapes of $\ase$ (bottom-right).
The ratio $B(91)/C(91)$ not belonging to the fit exhibits
strong $C_2$-dependence, contrary to generic $c/c$ and $b/b$ ratios.
}
\label{hqfig7}
\end{figure}

\paragraph{In Fig.\protect\ref{hqfig7}}
results of the $G_2$--model fit to the {\em ratios}\/ of $\lrang{x_Q}$
at different energies are shown.
Solid curves accumulate the squared errors in the description of four
ratios: $D^*(91)/D^*(10)$, $D^*(91)/D^*(30)$ and two ratios
from leptonic  quark  decays,
$C(91)/C(58)$ and $B(91)/B(32)$
As we have discussed above, one expects such ratios to be protected against
our ignorance about the confinement physics.
Indeed, a rather high stability in the quality of the fit inside a huge range
of variation of the ``confinement parameter'' $C_2$ is seen.
Bottom-right insertion displays corresponding shapes of $\ase$.
Within the chosen interval of $C_2$ the characteristic value of $\bar{A}$
(\ref{intcoups}) varies from $0.09$ ($C_2\!=\!10$, $\Lambda\!=\!0.7$)
up to $0.41$ ($C_2\!=\!1.4$, $\Lambda\!=\!0.5$)
that is changes by factor 2 in both directions
around the value 0.19 (\ref{intcoup}) that we have obtained describing
{\em absolute}\/ energy losses.

Also shown in Fig.\ref{hqfig7} is the mixed bottom-to-charm ratio
$B(91)/C(91)$.
This one does not belong here and, contrary to the
generic $c/c$ and $b/b$ quark ratios, strongly depends on $C_2$ as expected.

We observe that three of four generic ratios
show up no variation with $C_2$ at all.
It is the ratio $D^*(91)/D^*(10)$ only that induces some {\em negative}\/
correlation between $\bar{A}$ and $\Lambda$:
the latter moves to larger values with decrease of low-scale
interaction intensity.
Such a systematic drift is natural: charm quark mass is too small
to completely protect the normalization point $D^*(10)$
against $W$--{\em dependent}\/ (sic!)
confinement efects at total energy as low as $10\,\GeV$.
At the same time the first ratio dominates in the fit
while experimental accuracy of three others
is not sufficient at present to provide a direct safe way of measuring the
$\Lambda$ parameter.

Roughly one might present the result of fitting quark energy ratios as
\bminiG{ratress}\label{ratresL}
 \Lambda^{(5)} \>=\> 0.60 \>\pm\> 0.15\, (\mbox{stat.}) \>\pm\> 0.10\,
(\mbox{syst.})\>.
\eeeq
If statistical and systematic errors in Fig.\ref{hqfig7} were uncorrelated
(which is not the case), (\ref{ratresL}) would correspond to
\beeq\label{ratresA}
 \alpha_{\MSbar}(M_Z) \>=\> 0.128 \>\pm\> 0.007 \,.  
\emini
For the time being it will suffice to conclude that determination of
$\Lambda^{(5)}$ from quark energy ratios is consistent with that
from absolute energy losses, cf.  (\ref{loop2ress}).

\subsection{On bottom quark mass.}
\begin{figure}[htb]
\vspace{10 cm}
\includegraphics{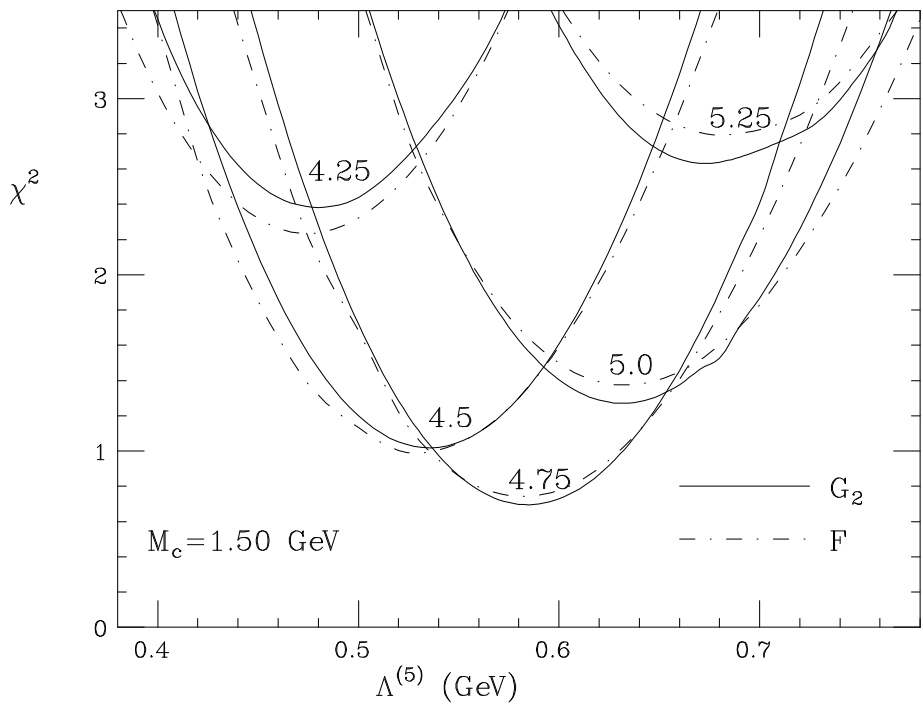}
\caption{ $M_b$ dependence. Fits within $F$ and $G_2$ models are shown
(two-loop, $n_f\!=\!5$.)
}\label{hqfig8}
\end{figure}

{Fig.\protect\ref{hqfig8}}
illustrates sensitivity of the total 7-fit to the bottom quark mass.
Here we have fixed $M_c=1.5\, \GeV$ and looked for minimal $\chi^2$ with
respect
to variation of $A/C_2$ (1-parameter fit) for given $\Lambda$ and $M_b$.

Quark mass dependence of the spectrum
(\ref{Ddef})--(\ref{radiator}) and, thus, of energy losses is basically
logarithmic (apart from non-relativistic corrections $(M/W)^2$ and
``confinement'' effects $\mu_\alpha/M$).
Nevertheless, a certain range of ``preferable'' bottom quark masses may be read
out from Fig.\ref{hqfig8}.
Within one standard deviation ($G_2$--model)
\beq
  M_b \>=\> 4.73 \> \pm\> 0.38 \> \GeV\,.
\eeq
A better understanding of the nature of the mass parameter $M$
as it enters PT formulae is needed before
such an analysis may be used for ``measuring'' bottom quark mass.

As for now, one finds satisfaction in noticing that the value $M_b=4.75 \GeV$
we have been using throughout this paper\refup{Voloshin}
does fit nicely into the PT picture of mean energy losses.

\subsection{Integrated coupling as invariant of the fit.}
Being free to play around with the detailed shape of $\ase$ in the origin,
one finds it necessary at the same time to fix some characteristic
measure of the radiation intensity in the non-PT momentum region.
Fig.\ref{hqfig5} suggests trying an {\em area}\/ under the curve
for an invariant parameter of the fit.
The result is quite impressive: areas under the $G_{2-4}$ and $F$
couplings are practically indistinguishable.
Introducing
\bminiG{intcoups}\label{intcoupdef}
\bar{A}(\mu) \>\equiv\> \frac1{\mu} \int_0^{\mu} dk\>\frac{\ase(k)}{\pi}\>,
\eeeq
we get (with one standard deviation error estimated from the $G_2$--model
margin, see Fig.\ref{hqfig5})
\beeq\label{intcoup}
 \bar{A}(2\,\GeV) \>=\>  0.190 \pm 0.010  \>.
\emini

\noindent
This integral measure not only proves to be quite stable
against the choice of {\em low}\/-momentum regularization but also
resistant to different approximations for the {\em high}\/-momentum tail of
$\ase(k)$.

\begin{table}[htb]
\begin{center}
\begin{tabular}{  c | c@{=}r | c|  c  | c || c }
 {Model for} & \multicolumn{2}{|c|}{Best-fit} & & &
$\chi^2$ & Integral $\bar{A}$\\
{$\ase$} & \multicolumn{2}{|c|}{parameter}
& \raise 0.6 em \hbox{$\Lambda \>(\MeV)$}
& \raise 0.6 em \hbox{$\alpha_{\MSbar}(M_Z)$}
& 5 d.o.f. & (\ref{intcoup}) \\[2 mm]
 \hline\hline
2--loop,    & $A$   & 0.184 & 730 $\pm$ 95  & 0.125  & 0.66  & 0.184 \\
with        & $C_2$ & 2.48  & 725 $\pm$ 85  & 0.125  & 0.64  & 0.191 \\
$c$, $b$    & $C_3$ & 4.41  & 710 $\pm$ 80  & 0.124  & 0.66  & 0.193 \\
thresholds; & $C_4$ & 7.79  & 705 $\pm$ 80  & 0.124  & 0.68  & 0.194\\
\cline{2-7}
$\Lambda^{(3+2)}$  & $C_1$  & 1.38 & 820 $\pm 100$ & 0.128 & 0.69 & 0.189 \\
\hline
2-loop,   &  A & 0.190     & 585 $\pm$ 85  & 0.127   & 0.73  & 0.187 \\
with 5    & $C_2$ & 2.88   & 585 $\pm$ 80  & 0.127   & 0.68  & 0.191 \\
massless  & $C_3$ & 5.59   & 575 $\pm$ 80  & 0.127   & 0.69  & 0.193 \\
quarks;   & $C_4$ & 10.8   & 575 $\pm$ 75  & 0.127   & 0.70  & 0.193 \\
\cline{2-7}
$\Lambda^{(5)}$ & $C_1$ & 1.46 & 655 $\pm$ 95 & 0.130 & 0.75  & 0.190 \\
\hline\hline
1-loop,   &     A & 0.193  &  480 $\pm$ 70 & 0.120  &0.71 & 0.188 \\
with 3    & $C_2$ & 2.22   &  485 $\pm$ 65 & 0.120  &0.66 & 0.190 \\
massless  & $C_3$ & 3.94   &  480 $\pm$ 60 & 0.120  &0.66 & 0.191 \\
quarks;   & $C_4$ & 6.90   &  480 $\pm$ 60 & 0.120  &0.66 & 0.191 \\
\cline{2-7}
$\Lambda^{(3)}_{\mbox{\small 1-loop}}$& $C_1$& 1.20& 515 $\pm$ {75}
& 0.122& 0.69 & 0.189 \\ \hline
1-loop,  &  A     & 0.209  & 310 $\pm$ 50   & 0.133   & 0.90  & 0.189 \\
with 5   & $C_2$ & 2.68    & 315 $\pm$ 50   & 0.133   & 0.79  & 0.191 \\
massless & $C_3$ & 5.48    & 315 $\pm$ 50   & 0.133   & 0.78  & 0.191 \\
quarks;  & $C_4$ & 10.9    & 315 $\pm$ 50   & 0.133   & 0.79  & 0.191 \\
\cline{2-7}
$\Lambda^{(5)}_{\mbox{\small 1-loop}}$& $C_1$& 1.28& 330 $\pm$ 50
& 0.134& 0.84& 0.189\\
\hline
\end{tabular}
\end{center}
\caption{Best 7--fits within various models for effective coupling.
$A/C$, $\alpha_{\MSbar}$, $\chi^2$ and $\bar{A}$ are given for the
central $\Lambda$ values.}
\label{tab2}
\end{table}

Table~2 accumulates charateristics of various 7-fits and demonstrates
an amusing stability of the value (\ref{intcoup}).
This justifies the qualitative expectation
of subsection~3.1 that it is the integral of the coupling
as a characteristic measure of confinement
(hadronization)  effects in inclusive energy spectra,
$\mu_\alpha$ of eq.(\ref{mualdef}),
that is responsible for the low-momentum contribution
to mean energy losses.

Thus we find empirically that the characteristic integral (\ref{intcoups})
turns out to be a {\em fit-invariant}\/ quantity
which one has to keep fixed to describe the absolute values of energy losses.
As pointed out by V.N.~Gribov,
it can be looked upon
as the long-distance contribution to the QCD field energy of a heavy quark.
It is worthwhile to notice that such an integral appears in the relation
between the running heavy quark mass at scale $\mu$
and the pole mass\refup{banda}
\beq\label{bandaeq}
 M^{\mbox{\small pole}} - M(\mu)
= \frac{8\pi}{3} \int_{\abs{\vec{k}}<\mu} \frac{d^3k}{(2\pi)^3}
\frac{\as(k)}{{k}^2} \> =\> C_F \int_0^\mu
d\kappa\> \frac{\as(\kappa)}{\pi}  \equiv
C_F \mu_\alpha\>.
\eeq

\subsection{Two-loop $\ase$ with heavy quark thresholds and
$\protect\alpha_{\MSbar}$ determination.}
\begin{figure}[htb]
\vspace{10 cm}
\includegraphics{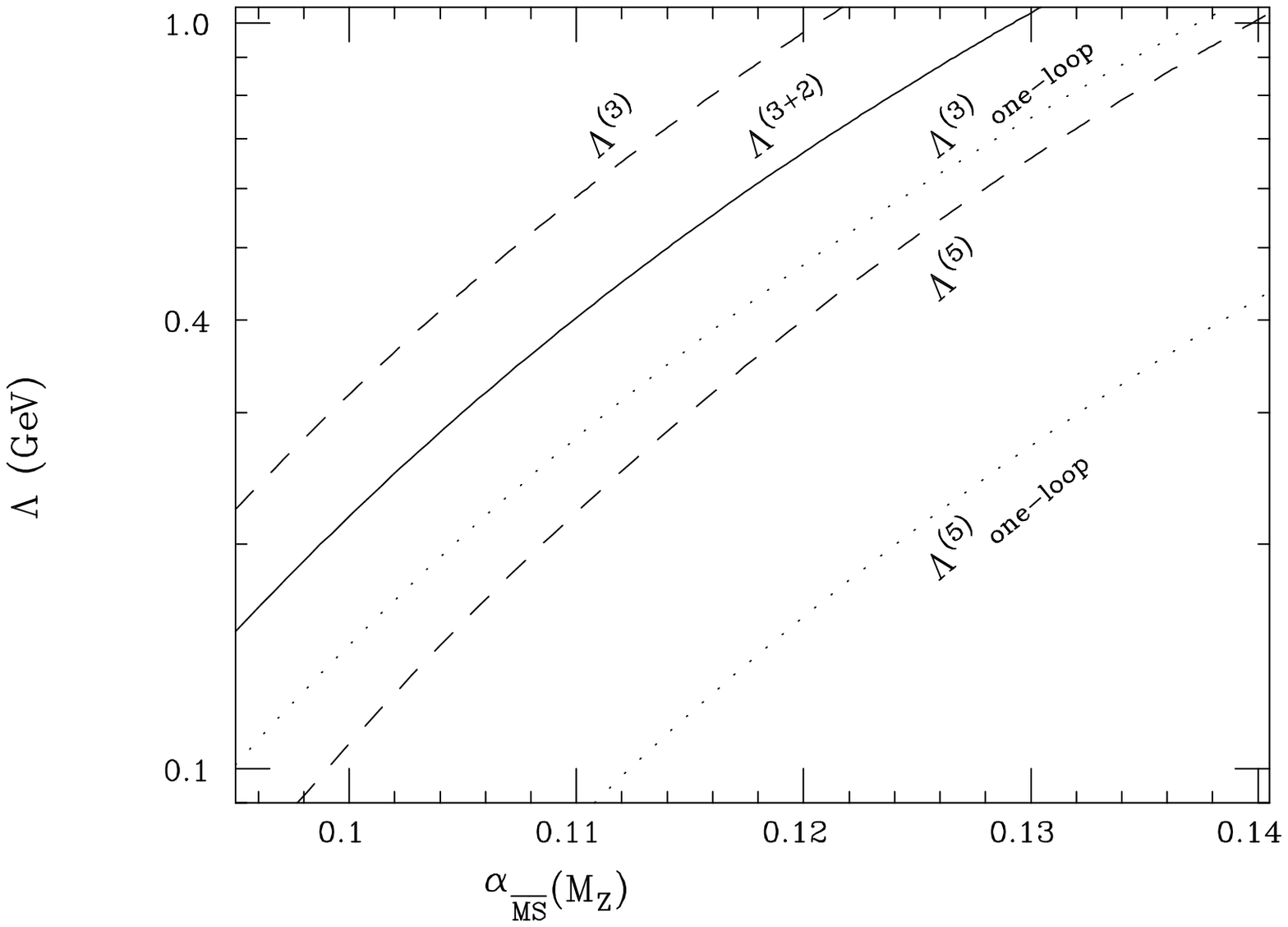}
\caption{Relation between $\Lambda$ and $\alpha_{\MSbar}(M_Z)$:
one-loop (dotted), two-loop (dashed) for 3 and 5 massless quarks
and the two-loop coupling with c, b quark thresholds (solid).
}\label{hqfig9}
\end{figure}

It is important to notice that the different approximations
for the high-momentum tail of $\ase$ listed in Table~\ref{tab2}
provide similar quality fits and preserve the $\bar{A}$ value
but, at the same time, lead to systematically different
values of $\alpha_{\MSbar}(M_Z)$.
{Fig.\protect\ref{hqfig9}}
helps to relate the values of $\Lambda$ parameter for different approximations
for the running $\ase$ to the reference value $\alpha_{\MSbar}(M_Z)$.

In the problem under consideration
one probes the coupling at $M_Z$ scale only indirectly.
For the first thing, ``half'' of the data belong to smaller $W$s.
Moreover, even when the LEP measurements are concerned,
the main contribution to energy losses originates from a broad
logarithmic integral of $\ase(k)$ running from
$k\la M_Q$ up to $k\la M_Z$, so that momenta just above $M_b$
(as well as between $M_c$ and $M_b$) play quite an essential role.
The reference values of $\alpha_{\MSbar}$ appearing in Table~\ref{tab2}
emerge as a result of extrapolation from intermediate
momentum scales that dominate in the fit.
Such an extrapolation is sensitive to details of high-momentum behaviour
of the running coupling.  When $n_f$ is taken smaller and/or
the two-loop effects are being taken into account,
$\ase$ becomes a steeper falling function of momentum and
the resulting value of $\alpha_{\MSbar}(M_Z)$  decreases.

\begin{figure}[htb]
\begin{minipage}{3.0 in}
\vspace{9 cm}
\includegraphics{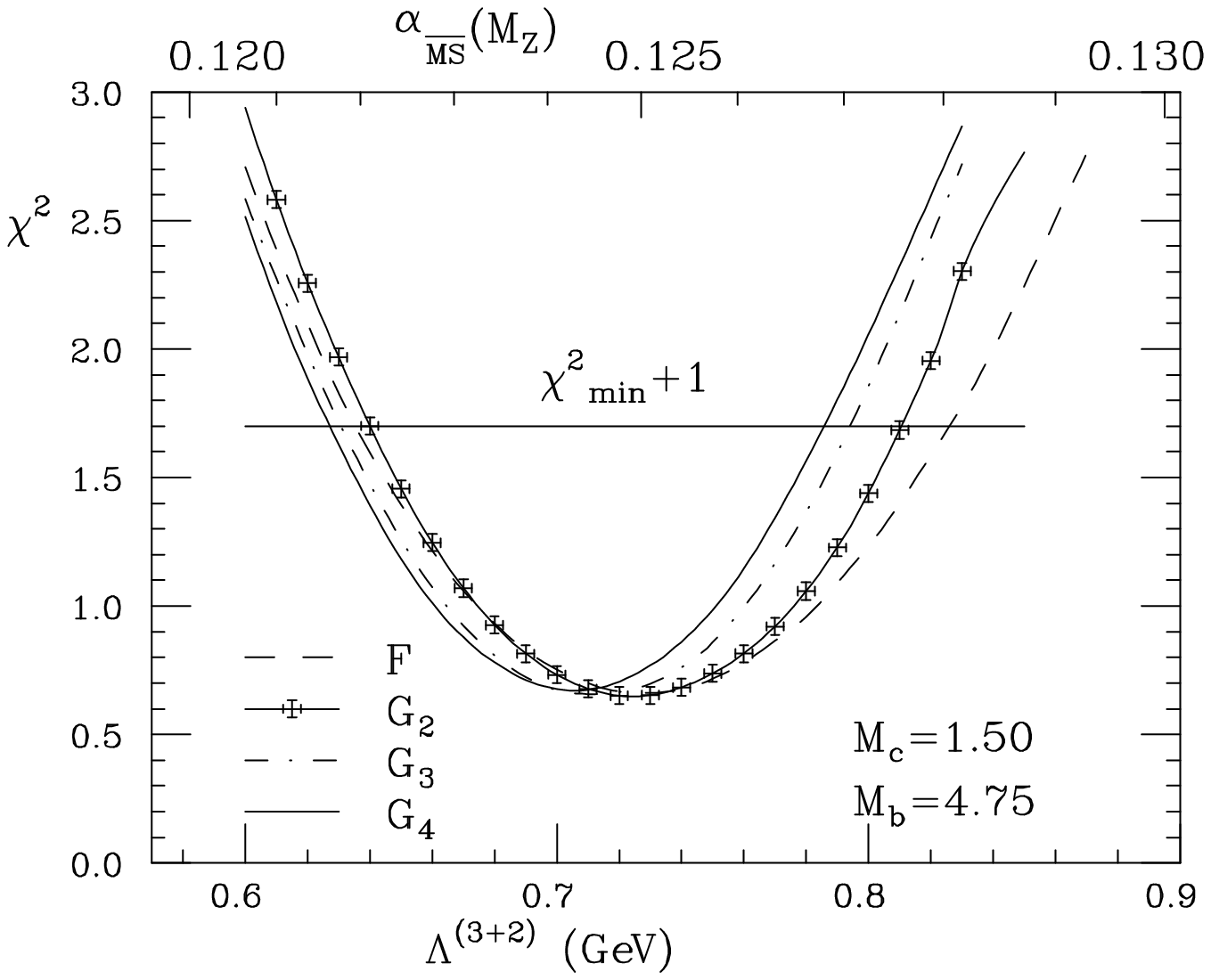}
\caption{ Quality of the total fits versus $\Lambda$
(two-loop $\ase$ with c, b thresholds).}
\label{hqfig10}
\end{minipage}
\quad
\begin{minipage}{3.4 in}
\vspace{9 cm}
\includegraphics{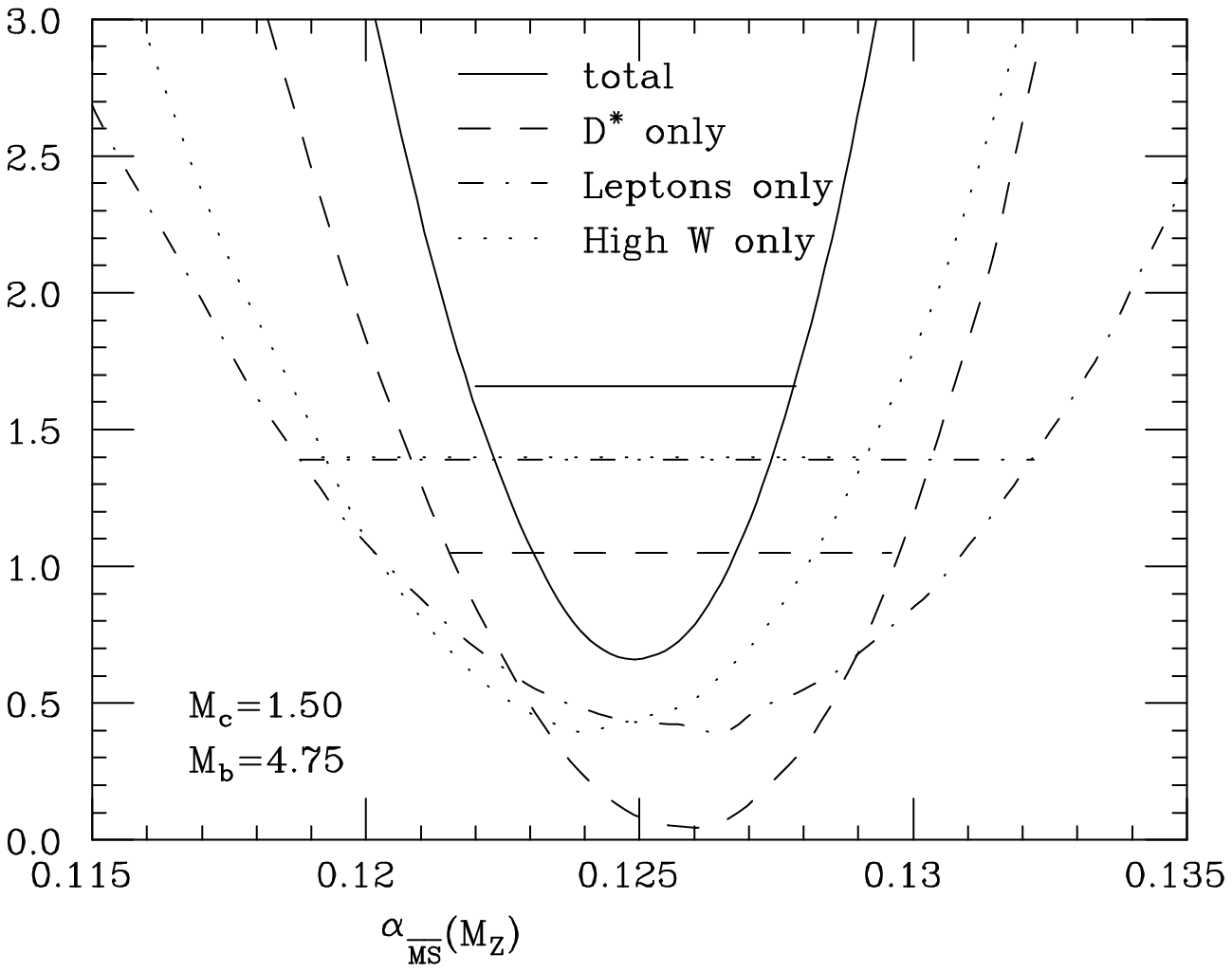}
\caption{On consistency of $G_2$--model fits to subsets of data
(two-loop; c, b thresholds).}
\label{hqfig11}
\end{minipage}
\end{figure}

Even an account of heavy quark thresholds
(which makes $\ase(k)$ a steeper function below $k\!\la\!M_b$)
drives down the $\alpha_{\MSbar}$ value.
To demonstrate this effect we include Figs.\ref{hqfig10} and \ref{hqfig11}
showing quality of the PT description of absolute energy losses
with account of the second loop {\em and}\/ c,~b quark threshold effects
in the running coupling.
In Fig.\ref{hqfig10} model dependence of the total 7--fit is shown
(cf. Fig.\ref{hqfig4});
Fig.\ref{hqfig11} collects fits to different subsets of data
(cf. Fig.\ref{hqfig6}).

Our conclusions about relative stability against the choice of
low-momentum regularization model and about consistent description of
leptonic, $D^*$ and high-$W$ data hold.
{}From these plots one obtains (see also Table~\ref{tab2})
\bminiG{threshress}
 \label{threshresL}
 \Lambda^{(3+2)} &=& 0.720 \>\pm\> 0.080\,(\mbox{stat})\>\pm\>
0.015\,(\mbox{syst})
\> \GeV\>,
\eeeq
where we have singled out systematic uncertainty due to model dependence.
This results in
\beeq
 \label{threshresA}
 \alpha_{\MSbar}(M_Z) &=&  0.125 \>\pm\> 0.003 \,.
\emini
It is important to stress that the second loop effects
are there in the running coupling and $n_f$ is not a free parameter:
quarks (and their masses) are what they are.
Therefore a huge interval of $\alpha_{\MSbar}$
values seen in Table~\ref{tab2} has nothing to do with actual
theoretical uncertainty in determining this important datum.

Before we turn to discussion of systematic errors
some comment is in order.
It has to do with the number of acting quark flavours and will eventually
give us an additional consistency check of the approach.

An attentive reader could have noticed that
using 5--flavour running coupling seems to contradict
the very logic of the present paper.
Indeed, in Section~\ref{PTsec} we constructed the energy distribution
corresponding to {\em direct}\/ production of a heavy quark $Q$.
To do so we disregarded sea mechanism of heavy quark
production and omitted the second loop anomalous dimension term
related to $\QQ\QQ$ final states.

However, there is yet another subtle source of copious $\QQ$ pairs,
namely, the running coupling $\ase$ embodied into PT radiator.
The way the coupling runs in theoretical expressions for inclusive
characteristics depends on experimental setup:
Veto on fermion pair production in the final state
suppresses effective interaction strength at large momentum scales
(QCD coupling decreases faster; an increase of $\alpha_{em}$ slows down).
Therefore, to preserve the logic of the approach, we better make it clear
that the use of $n_f\!=\!3$ effective coupling
which goes along with suppression of additional $c\bar{c}$, $b\bar{b}$ pairs
is consistent with the reported result (\ref{threshresA}).

\begin{figure}[htb]
\vspace{11.5cm}
\includegraphics{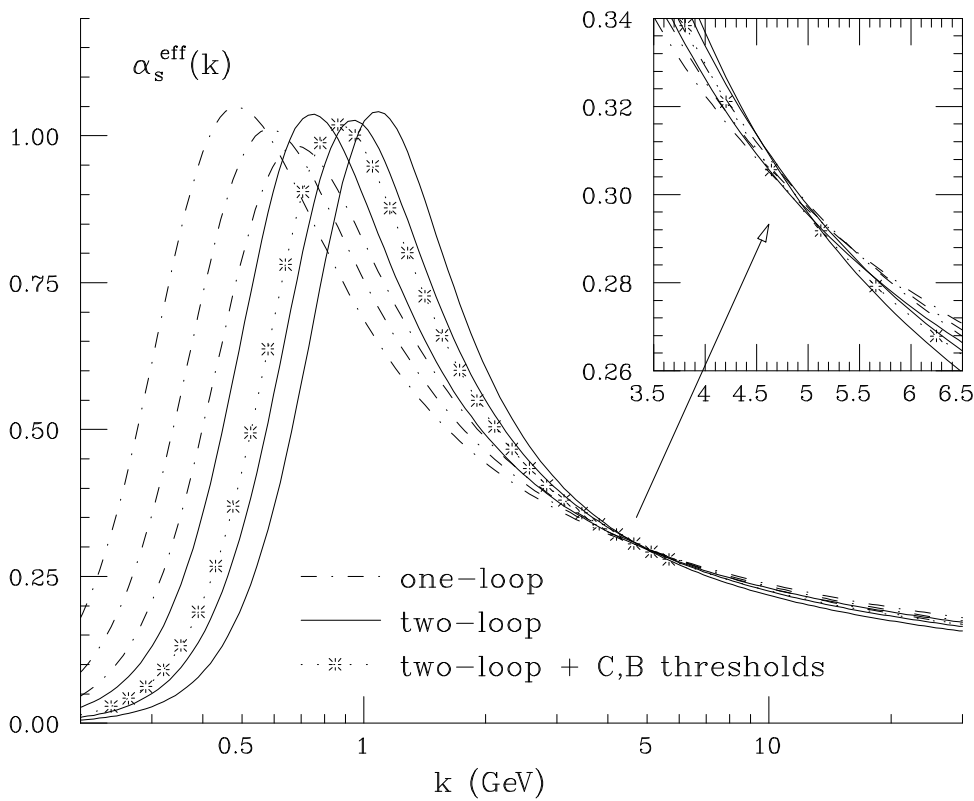}
\caption{Best-fit $G_2$--model effective couplings.
$n_f=5,4,3$ progress from left to right.}\label{hqfig12}
\end{figure}

In Fig.\ref{hqfig12} $G_2$--model couplings are shown
with the best-fit parameters $\Lambda$ and $C_2$ listed in Table~\ref{tab3}.
As far as low momenta are concerned, variations due to $n_f$ and one vs.
two loops should be looked upon simply as different models for trial
regularized coupling.
All of them do the job.
In particular, the 3--flavour $\ase$ which interests us at the moment
(the last line of Table~\ref{tab3}) does provide an excellent fit,
$\chi^2$/d.o.f.$\approx 0.6/5$.

\begin{table}[htb]
\begin{center}
\begin{tabular}{  c | c | c|  c  | c | c | c||c|| c | c }
 $G_2$--model & $n_f$ & $\Lambda^{(n_f)}$ & $C_2$ & $\chi^2$ & $\bar{A}(2\GeV)$
& $\ase(2\GeV)$ & $\ase(\!M_b\!)$ & $\ase(\!M_Z\!)$ &
$\alpha_{\MSbar}(\!M_Z\!)$
 \\[1 mm]  \hline\hline
+thresholds & 3+2 & 725 & 2.48 & 0.65 & 0.191 & 0.530 & 0.303 & 0.135 & 0.125\\
\hline
        & 5 & 315 & 2.68 & 0.79 & 0.191 & 0.443 & 0.302 & 0.145 & 0.133 \\
1--loop & 4 & 395 & 2.43 & 0.72 & 0.191 & 0.463 & 0.303 & 0.139 & 0.126 \\
        & 3 & 485 & 2.22 & 0.66 & 0.190 & 0.488 & 0.306 & 0.134 & 0.120\\
\hline
        & 5 & 585 & 2.88 & 0.69 & 0.191 & 0.494 & 0.303 & 0.138 & 0.128\\
2--loop & 4 & 770 & 2.72 & 0.64 & 0.190 & 0.549 & 0.305 & 0.131 & 0.120\\
\cline{2-9}
        & 3 & 935 & 2.56 & 0.62 & 0.190 & 0.608 & 0.305 & 0.124 & 0.112 \\
\cline{2-9}
\end{tabular}
\end{center}
\caption{\label{tab3} Characteristics of the best $G_2$--model fits.}
\end{table}

High stability of the $\bar{A}$ value (\ref{intcoup}) is confirmed once again.
Moreover, one more interesting fit-invariant quantity emerges,
namely, the value of the effective coupling at the bottom mass
scale\footnote{since magnitude of $\ase$ around $5\GeV\gg\Lambda$ is
insensitive
to $A/C$ and depends only on $\Lambda$, the same conclusion (\ref{asbott})
follows from $F$ and $G_{3,4}$ models},
\beq\label{asbott}
\ase(M_b) \>\approx\>  0.30 \,.
\eeq
The curves in Fig.\ref{hqfig12} are quite different below 2\GeV,
focus around 5\GeV and diverge again at high momenta.
The columns adjacent to the double-lined one in Table~\ref{tab3}
illustrate this behaviour.
In particular, the 3--flavour coupling continued to the LEP scale is
substantially lower compared to the ``3+2'' coupling we have been using
before (0.124 vs 0.135).
However, it would be erroneous to read out
$\alpha_{\MSbar}=0.112$ from the last line of Table~\ref{tab3}.

As we see, heavy flavours in $\ase$ are indeed
irrelevant for the problem under consideration.
However, they do contribute to the evolution of the standard QCD coupling.
Therefore, having found the magnitude of $\ase$ at intermediate scales
$k\sim 5\GeV$ which dominate in the problem,
one has to apply the 5--flavour evolution from $\sim5$ up to 91 \GeV\
aiming at extraction of the
reference value $\alpha_{\MSbar}(M_Z)$.

Since the starting value
$\ase(M_b)=0.305$ practically coincides with that from the
previous ``3+2'' analysis (0.303), so do the results.
Choosing the starting point of high momentum extrapolation in between
$k=M_b/2$ and $k=2M_b$ leads to
\beq
  \alpha_{\MSbar}(M_Z) \>=\>
    0.125 \>\pm\> 0.003\,.
\eeq
The, slightly inflated, error here is moderate,
and the central value perfectly matches
with the result of the previous analysis (\ref{threshresA})
based on ``3+2'' coupling.

\subsection{Systematic errors and prospects.}
First of all, there is a problem of correspondence between theoretical
prediction and the data. As we have stressed above, formulae of this
paper were designed to describe mean energy of a primary quark produced
in the \ee annihilation vertex.
We were treating experimental numbers as corresponding to direct \QQ
production.
We do not feel in a position to judge to what extent the experiments
actually met such an expectation.

The good news is however that the bottom sector is safe in this respect:
$g\to b$ sea component is vahishingly small at present,
and so is charm production at low energies\refup{MikeS}.
Therefore only LEP charm data should concern us here.
Bearing this is mind it is important to notice that the $D^*$ LEP datum \#3
(which is the most precise measurement and therefore practically the only one
vulnerable)
does correspond to primary charm\footnote{An older average
$\lrang{x_{D^*}}= 0.494\pm 0.014$ which we have used in the previous
analysis\refup{preprint} was contaminated by the sea.
We are indebted to P.M\"attig for clarifying this point}.
MC modeling was used to subtract the gluon component (with an uncertainty
included into experimental systematic error).
Reliability of such a subtraction has been recently verified by the first
OPAL measurement of the charm production via gluons\refup{PeterCC}.

Apart from this,
one can think of the following sources of systematic error
of $\alpha_{\MSbar}(M_Z)$ determination.

\vspace{3 mm}
{\em higher orders.}\quad {\bf Estimate:} 0.002; {\bf Comment:} optimistic.\\
Follows from analysis of approximate relation (\ref{phco})
between physical coupling $\ase$ and $\alpha_{\MSbar}$
and of forceful exponentiation of $\as$ terms in the radiator (\ref{radiator}).
Apart from three-loop analysis,
an exact $\cO{\as^2}$ theoretical calculation of $\lrang{x_Q}$ would
be helpful.

\vspace{3 mm}
{\em power effects in}\/ $\ase(k>2\GeV)$.\quad 0.003;
arbitrary (hopefully conservative). \\
Is based on comparison with the ``soft'' $G_1$--model (see above).
Variation within ``hard'' regularization models ($F$, $G_{2-4}$)
is below 0.001 (see (\ref{threshress})).
To gain quantitative theoretical control over the $1/k^2$ power term
in the effective coupling and, thus, to reduce corresponding systematic error,
might be not as hopeless a goal as it seems.

\vspace{3 mm}
{\em kinematical effects.}\quad  0.000; safe.\\
The structure of the PT spectrum is such that it respects the kinematical
boundary $x\ge x_{\min}=2M/W$ only in the first \as order.
Contribution to $\lrang{x_Q}$ from potentially dangerous region between
$x_{\min}$ and $2x_{\min}$ can be estimated as
$$
\Delta \lrang{x_Q} \sim {\as(W)}/{\pi}\cdot \left( 1- \lrang{x_Q}\right)
\cdot x_{\min}^2 \,,
$$
which value is well below 1\% even for $x_{\min}$ as big as 1/3.

\vspace{3 mm}
{\em quark masses.}\quad   0.002; conservative. \\
Given that the structure of low-momentum contribution
proportional to $\bar{A}$ naturally embodied into PT analysis
reminds that of the shift from Euclidean to ``on-shell'' quark mass
(see (\ref{bandaeq})),
one may worry about {\em double counting}\/
if the pole mass is used for $M$ in the PT formulae.
The problem should be studied theoretically.
Meanwhile, let us mention that one may get equally good description of
absolute values of mean quark energy losses
with {\em smaller}\/ (Euclidean?) quark masses plugged in.
For example, by taking
$$
 M_c= 1.30 \,\GeV\>; \quad M_b=4.50 \GeV
$$
one obtains\footnote{4.5 is a preferable bottom mass value for $M_c\!=\!1.3$,
analogously to $M_b\!=\!4.75$ for $M_c\!=\!1.5$; see above, Fig.\ref{hqfig8}}
\begin{center}
\begin{tabular}{l @{=} l | c| c| c| c }
\multicolumn{2}{ c|}{$G_2$--model} & $C_2$ & $\chi^2$ & $\bar{A}(2\GeV)$
& $\alpha_{\MSbar}(M_Z)$ \\[2mm] \hline
$\Lambda^{(5)}$     & 600 & 3.60 & 0.72 & 0.166 & 0.1280 \\
$\Lambda^{(3+2)}\>$ & 740 & 2.94 & 0.66 & 0.166 & 0.1257
\end{tabular}
\end{center}
We observe that the resulting value $\alpha_{\MSbar}(M_Z)$
hardly increases by 1 per mil.

Another feature of the alternative set of quark masses
worth noticing is a systematic decrease of the value of characteristic
integral $\bar{A}$.
This may be a welcome trend bearing in mind a recent analysis of power
corrections to jet shape variables, see \cite{shapes}.
For the time being we choose to look upon this 10\% shift as
a systematic uncertainty which will be greatly reduced when proper
theoretical understanding of the nature of the quark mass parameter is
achieved.

Having said that we modify (\ref{intcoup}) and present the final result
for the integrated coupling as
\beq\label{finAres}
\bar{A}(2\GeV)\>=\>0.18\>\pm\>0.01\,(\mbox{stat})\>\pm\>0.02\,(\mbox{syst})\,.
\eeq

\begin{figure}
\vspace{13.5cm}
\includegraphics{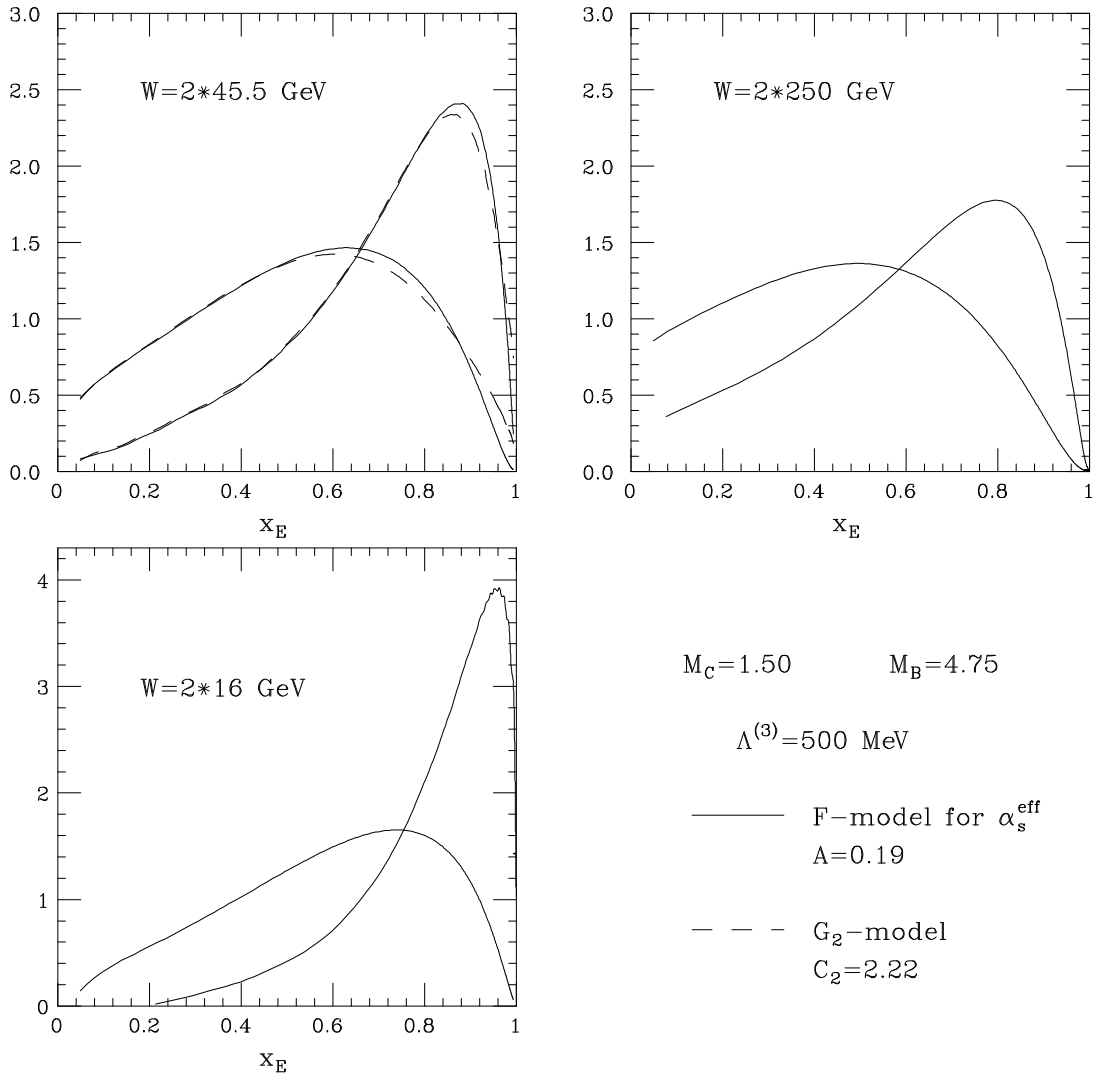}
\caption
{\label{hqfig13}
Evolution of inclusive c, b spectra from $W=32\GeV$ to $W=500\GeV$.
The best-fit $F$-model (solid) and $G_2$-model curves (dashed) are shown.}
\end{figure}

Finally, in Fig.\ref{hqfig13} we demonstrate the $W$-evolution of realistic
c- and b-quark spectra obtained within the $F$-model with one-loop 3-flavour
$\ase$.
These curves correspond to the values of $A$ and $\Lambda$ providing the
best common fit to mean energy losses as described above.
For comparison the best-fit $G_2$-model curves are also shown for LEP energy.
$F$- and $G$-model spectra are quite close to each other.
We may conclude that is suffices to fix the integral parameter (\ref{intcoup})
together with the value of $\Lambda$
to predict the differential energy distributions with a reasonable
accuracy\footnote{Let us notice that the differential quark energy
spectra obtained by the inverse Mellin transform (\ref{Ddef}), (\ref{Dj})
violate kinematical boundary $D(x\!<\!2M/W)\!=\!0$ at $\cO{\as^2}$ level}.

A stable numerical procedure for numerical evaluation of the inverse Mellin
transform (\ref{Ddef}) remains to be designed.

\mysection{Conclusions}
Jets initiated by heavy quarks (b,c) are now extensively studied
experimentally.
The interest to this subject is connected not only with testing the
fundamental aspects of QCD but also with its large potential importance
for measurements of heavy particle properties: lifetimes, spatial oscillations
of flavour, searching for CP violating effects in their decays \etc
Properties of b--initiated jets are of primary importance for analysis of the
final state structure in the \tt\ production processes.

In this paper we presented results of the study of the inclusive energy spectra
of leading heavy-flavoured particles ($H_Q$)
based on the perturbative expression (\ref{Ddef}) -- (\ref{radiator})
for heavy $Q$ distributions that emerge after taking into proper account
multiple gluon \br\ off the \QQ pair.

Our approximation includes the two-loop anomalous dimension,
keeps track of the {\em collinear}\/ logarithms $a\ln W$ and $a\ln M$,
{\em soft}\/ double-logarithmic $a\ln^2(1\!-\!x_Q)$ and essential
single-logarithmic $a\ln(1\!-\!x_Q)$ contributions in all orders.
At the same time, it embodies the exact first order result ${\cal{O}}(a)$
for the inclusive energy distribution which property is necessary to
account for non-relativistic suppression of gluon radiation in the situation
when jet energies are comparable with a heavy quark mass.

The spectra we have been considering correspond to {\em direct}\/ $Q$
production. Such an approach disregards the sea contribution to heavy quark
yield which is numerically small at present and foreseeable energies.
Accordingly, a specific term in the non-singlet two-loop anomalous
dimension has been omitted which is due to $\QQ\QQ$ final states and would
violate the ``multiplicity'' sum rule (\ref{sumrule}).

Apart from this simplification, the formal relative accuracy of the
perturbative result (\ref{Ddef}) -- (\ref{radiator}) is $\cO{a^2(W^2)}$ which
estimate is {\em uniform}\/ in $x_Q$.  Perturbative corrections at the quark
mass scale which were left out of control start from $\cO{a^3(M^2)}$.

We have formulated the perturbative result (\ref{Ddef}) -- (\ref{radiator})
in terms of a ``physical'' coupling, the one which directly measures radiation
intensity of relatively soft gluons\refup{CMW90}.
This coupling constant is different from (roughly, 10\% larger than)
$\alpha_{\MSbar}$ and is approximately related to the latter by (\ref{phco}).
Expressed in terms of $\ase$, the radiator (\ref{radiator})
acquires the two-loop contribution $\tilde{\Delta}^{(2)}$ which
is free from an artificial $n_f$-dependence\refup{BLM} and proves
to be numerically negligible.

There are two ingredients of the standard approach to description of $H_Q$
spectra. Here one starts from the phenomenological non-PT fragmentation
function for the $Q\to H_Q$ transition and then traces its evolution with
increase of the annihilation energy $W$ by means of PT QCD.
Being formally well justified for describing the $W$-evolution, this approach
however leaves the dependence on the quark mass, $M$, basically out of the PT
control.\footnote{A plausible rescaling procedure for extracting the non-PT
part of the $B$ meson fragmentation from the existing data on $D$ meson spectra
has been suggested recently in \cite{CoNa}.}

Motivated by the LPHD concept, we attempted to derive {\em pure}\/ PT
predictions without invoking the phenomenological fragmentation function.
Starting point for such an attempt was the observation that an appearance
of the parton model motivated {\em peak}\/ in the non-PT fragmentation at
large $x_Q$\refupd{leading}{fragm} can be attributed to the Sudakov
suppression effects {\em provided}\/ one feels courageous
enough to continue the PT description down to the region of gluon transverse
momenta, $\kp\sim \Lambda\cdot(W/\sqrt{M\Lambda})^{0.2}$,
which at present (and foreseeable) energies looks dangerously
close to the non-perturbative domain.

When getting rid of the transverse momentum cutoff one faces the problem
of the formal ``infrared pole'' in
$\as$ and is forced to introduce the effective
non-singular coupling $\ase(k)$ that remains finite at $k\!\le\!\Lambda$.
It is not easy to justify the very notion of $\as$ at small momenta where
the PT quark-gluon language seems to be hardly applicable at all.
In the problem under consideration it can be related to the effective measure
of intensity of accompanying particle production at the confinement stage of
the $H_Q$ formation, a finite number of light hadrons produced in addition to
the ($W$ dependent) particle yield due to PT-controlled gluon \br at the first
stage of the hard \QQ pair creation process.

We have checked that, in accordance with expectations based on general
factorization properties, our ignorance about confinement does not affect
the $W$-dependence of the mean energy losses $\lrang{x_Q}(W,M)$.
Thus the ratio $\lrang{x_Q}(W)/\lrang{x_Q}(W_0)$
can be used to extract the fundamental QCD parameter $\Lambda$.
Another infrared stable quantity found empirically is the PT prediction for
the normalized peak position, ${x^{peak}_Q}(W)/{x^{peak}_Q}(W_0)$.

At the same time, the absolute values of $\lrang{x_Q}$, peak positions and
particle distributions in general show up substantial variations with
the value (momentum dependence) of $\ase$ in the origin chosen as an
input for calculating the PT {\em radiator}\/ (\ref{radiator}).
We have compared experimental values of $\lrang{x_Q}$ extracted from
leptonic c- and b-quark decays and $\lrang{x_{D^*}}$
directly with the PT motivated prediction for the quark energy losses.

Our treatment of the hadronization stage that implicitly appeals to
duality arguments makes it plausible to apply purely perturbative approach
to inclusive quantities such as the lepton energy distributions.
Meanwhile, it can not pretend to describe {\em exclusive}\/ heavy hadron
spectra such as those of $D^*$, for example.
Therefore, taking the measured mean energies of $D^*$
for $\lrang{x_Q}$ might introduce some systematic uncertainty to the fit,
the error which is difficult to estimate.

Nonetheless, such a comparison has demonstrated that radiative charm and beauty
losses may be consistently described in the energy range $W=10\to 90\GeV$
within a variety of models for low momentum behaviour of the infrared finite
effective coupling $\ase$.  By tuning the shape parameter of a model
one achieves an accuracy of the common fit
as good as $\chi^2$/d.o.f.$\approx 0.7/5$
(7 data points; two free parameters, $\Lambda$ and $A/C$).

We have checked that the fit to all data is consistent with the fits to
high-$W$, $D^*$ and inclusive lepton data samples.

{}From the first sight, we have gained not much profit since we had to pay
for eliminating an arbitrary non-PT fragmentation function by introducing
another arbitrary function $\ase(k)$ at $k\la 2\GeV$.
An essential physically important difference between two
approaches is that $\ase$ as a key ingredient of the PT-LPHD approach
is supposed to be {\em universal}\/ so that (quite substantial) differences
between inclusive spectra of c- and b- flavoured hadrons should be under
control.  By simultaneous fitting of the available experimental data on the
energy losses in charm and beauty sectors
we have demonstrated consistency of the hypothesis of $\ase$ universality.

Moreover, the same notion of the infrared-finite effective coupling
can be tried for a good many interesting problems in the light quark sector.
An incomplete list of such phenomena for which the PT analysis has been
carried out recently to next-to-leading order includes
transverse\refup{CTW} and longitudinal momentum distributions\refup{CTS}
in hadron-initiated processes, the energy-energy correlation\refup{KT},
the thrust\refup{CTWT} and the heavy jet mass distribution\refup{hjm}
in \ee annihilation.

The presence of the exponential of the characteristic integral over gluon
momenta which emerges after all-order resummation of the Sudakov
logarithms\refupd{pion}{PP} is a common feature of the corresponding PT
expressions.
It is straightforward to derive quantitative PT-motivated predictions
by implementing the universal $\ase$ in these integrals and, at the
same time, by getting rid of non-PT ``hadronization'' effects which are
usually taken into account by convoluting the PT distributions with
phenomenological fragmentation functions (initial parton distributions
for the case of hadron-initiated processes).

Carrying out this laborious but promising program one should get a valuable
information about the confinement physics as seen through the eyes
of the integrated influence of the large-distance hadron production upon
inclusive particle distributions and/or event characteristics.

To this end the numerical results of our analysis of the heavy quark energy
losses, namely,
\bminiG{results}
\label{alres}
(2\GeV)^{-1}\int_0^{2\GeV} \!dk\,\frac{\ase(k)}{\pi}\>\>&=&\>\>
0.18\pm 0.01 \, (\mbox{exp}) \>\pm\> 0.02\, (\mbox{theor}) \>,\\
\label{lamres}
\alpha_{\MSbar}(M_Z) \>\> &=& \>\> 0.125 \pm 0.003\, (\mbox{exp}) \>
                                      \> \pm 0.004\, (\mbox{theor}) \>,
\emini
should be looked upon as a first hint for the more detailed study including
the light quark phenomena discussed above.

One comment is in order concerning the nature of the results (\ref{results}).
The parallel description of c,b losses is sensitive to $\ase$ in the
low momentum range, say, of the order of (and below) $M_b$,
so that the integral characteristic (\ref{alres})
gets fixed quite sharply (theoretical error is due to
our ignorance about the heavy quark masses and should be eliminated in the
future).

At the same time such a description proves to be quite liberal to details of
the high-momentum behaviour of the coupling (one- versus two-loop $\as$,
number of active quark flavours).
{}From this point of view the value (\ref{lamres}) that one extracts in the end
of the day is rather a tribute to (unfortunate) tradition.
$\alpha_{\MSbar}(M_Z)$ should be looked upon as a reference value
which emerges after theoretical extrapolation rather than a quantity that is
``measured'' by the above analysis.
Theoretical error in (\ref{lamres}) is dominated by potential
$k^{-2}$ power correction effects in $\ase(k)$.

Let us stress again that in the low momentum region
behaviour of the effective coupling is poorly known not because
of the limited knowledge of higher order effects but rather because
of an essentially different physical phenomenon that enters the game,
the one that is usually referred to as confinement.

{}From this point of view our notion of the infrared finite $\ase$ differs
from one that emerges from the three-loop analysis based upon the
renormalization-scheme-invariant approaches to the \ee annihilation cross
section\refupd{CKL}{MStev} and the $\tau$-lepton hadronic
width\refup{CKL}.
Nevertheless it is worth mentioning that the value of the {\em couplant}\/
$\alpi(0)$ and the integral measure obtained in \cite{MStev}
in the framework of the Minimal Sensitivity Principle\refup{PMS}
are consistent with (\ref{alres}).

Phenomenological verification of the fact
that the effective QCD coupling stays numerically small would be of
large practical value.
Gribov theory of confinement\refup{BH} demonstrates how
colour confinement can be achieved in a field theory of light
fermions interacting with comparatively small effective coupling
--- a fact of potentially great impact for
enlarging the domain of applicability of perturbative ideology to
the physics of hadrons and their interactions.

\vspace {1 cm}
\noindent {\bf\large Acknowledgments }
\vspace {0.3 cm}

We are indebted to V.N.~Gribov and P.~M\"attig for many
helpful discussions and valuable comments.
One of us (Y.D.) is most grateful to the Theory Department of the University
of Lund for the warm hospitality extended to him during the years when this
study was being performed.
This work was supported in part by the UK Particle Physics and
Astronomy Research Council.

\newpage
\def\cS{{\cal{S}}}
\def\cZ{{\cal{Z}}}
\def\cL{{\cal{L}}}
\def\cH{{\cal{H}}}
\def\k1k2{\left(\frac{1\!-\!z_1}{1\!-\!z_2}+\frac{1\!-\!z_2}{1\!-\!z_1}\right)}

\bAPP{A}{Running coupling in the Radiator}

\section{Notation and Kinematics}
We consider production of a \QQ pair with on-mass-shell 4-momenta $p_1^\mu$,
$p_2^\mu$ and a gluon with momentum $k^\mu$ by a colourless current with
the total momentum $q^\mu$.
Hereafter for a sake of simplicity we measure all the momenta and the
quark mass in units of the total annihilation energy ($q^2\!\equiv\! W^2=1$).

In terms of quark and gluon energy fractions,
\baeq\label{kin}\eqalign{
z_i\equiv 2(p_iq)\,,\>\> z\equiv 2(kq) \,; & \qquad
 z_1+z_2+z  =2 \>; \cr
 2p_1p_2 &= (q-k)^2-2m^2 = 1-z\,-\, 2m^2 +k^2 = z_1+z_2-1\,-\,2m^2 +k^2 \>;
}\eaeq
virtual quark propagators are
\baeq\label{qprop}
 \kappa_1 \equiv (p_1+k)^2 -m^2= 1-z_2 \>, \quad
 \kappa_2 \equiv (p_2+k)^2 -m^2= 1-z_1  \>.
\eaeq
(Relations (\ref{qprop}) do not imply $k^2\!=\!0$.)

Three-body kinematics restricts the difference of quark energy
fractions as
\bminiG{delkin}
\label{z1z2dif}
 4 \,{|{\vec{k}}|^2}
\cdot \beta^2  \>\ge\> (z_1-z_2)^2  \>,
\eeeq
where $\beta=\beta(z)$ represents the quark velocity in the rest frame
of the \QQ pair (\QQ cms):
\beeq\label{betadef}
\beta^2 \equiv \frac{p^2_{\QQ}}{E^2_{\QQ}}
=   1- \frac{4m^2}{ (p_1+p_2) ^2}
=  1- \frac{4m^2}{ (q-k) ^2}  .
\emini

\subsection{$k^2=0$}
For the on-mass-shell gluon case one has
$4|{\vec{k}}|^2=z^2$; $(q-k)^2=1-z$
and (\ref{delkin}) gives
\baeq\label{beta0}
 \beta = \left(1-\frac{4m^2}{1-z}\right)^{\half} \>\>
 \ge \>\> \left|\frac{z_1-z_2}{z}\right|\,.
\eaeq

\paragraph{$z,\Theta_c$ Basis.}
When the gluon energy is kept fixed, it is convenient to use the
{\em scaled difference}\/ of quark energies
as a complementary variable
related to the gluon angle in the \QQ cms:
\baeq \label{udef}\eqalign{
u & \equiv  \frac{z_1-z_2}{ z}=  \frac{2(p_1-p_2)k}{2(p_1+p_2)k}
=  \beta \cos \Theta_c \>; \quad -\beta\>\le\> u\>\le\> +\beta \>.
}\eaeq
The maximal value $\abs{u}\!=\!\beta$
is reached when the gluon is {\em collinear}\/ with one of the quarks.

Introducing maximal quark velocity $v$,
\baeq\label{vdef}
\beta^2\le v^2 \>\equiv\>  1-4m^2 \> \ge\> z\>,
\eaeq
one may present the integration phase space as
\baeq\label{zuphsp}
 \int\> dz_1\>\int\>dz_2 \>\>=\>\> \half \int_0^{v^2}\> zdz\>
\int_{-\beta}^{\beta}\,du \>.
\eaeq
Useful relations:
\bminiG{usrel}
1- u^2 &=& 4 {(1-z_1)(1-z_2)}\cdot{ z^{-2}} \>; \\
 \k1k2 &=& \frac{z^2}{(1-z_1)(1-z_2)} -2 \>=\> \frac{4}{1-u^2} -2\>.
\emini

\paragraph{$x,y$ Basis.}
For the purpose of deriving single-inclusive quark spectrum
we break the symmetry between $z_1$, $z_2$ and denote
$$
 x \equiv z_1 \>;\>\>\> y\equiv 1-z_2\>.
$$
One has to fix $x$ and integrate over $y$
in the limits $ y_-\le y \le y_+$ which follow from the kinematical
restriction (\ref{beta0}).
In terms of $x,y$ the latter takes the form
\baeq
 y^2(1-x+m^2) - y(x-2m^2)(1-x) + m^2(1-x)^2 \> \le 0\,.
\eaeq
This gives
$$
 y_\pm = \frac{1-x}{2( 1-x+m^2)}  \left( x-2m^2 \pm \sqrt{x^2-4m^2} \right) .
$$
Introducing
\baeq\label{z0def}
z_0  \equiv \frac12  \left( x-2m^2 + \sqrt{x^2-4m^2} \right) =x + \cO{m^2}\,,
\eaeq
one may write (\cf (\ref{twoscales}))
\baeq\label{ymp}\eqal2{
 y_+ =&  \frac{(1-x)\, z_0}{ 1-x+m^2} \>\> &=\>\>
x \cdot\left(1+ \cO{m^2}\right) \>, \cr
 y_- =& \frac{1-x}{z_0} m^2 \>\> &=\>\>
\frac{1-x}{x} m^2 \cdot \left(1+ \cO{m^2}\right)\>.
}\eaeq
Useful relations:
\bminiG{usrel2}
y_+ - y_- &=&  \frac{1-x}{ 1-x+m^2} \,\sqrt{x^2-4m^2} \>, \\
(y_-)^{-1} - (y_+)^{-1} &=&  \frac{\sqrt{x^2-4m^2}}{ 1-x} \,m^{-2} \>, \\
(y_+)^2 - (y_-)^2 &=&  (x-2m^2)  \sqrt{x^2-4m^2}
\left(\frac{1-x}{ 1-x+m^2} \right)^2 .
\emini

\subsection{$k^2>0$}
If we allow the gluon to have a positive virtuality $k^2$, (\ref{delkin})
takes the form
\baeq
 (z_1-z_2)^2 \le (z^2-4k^2) \left( 1- \frac{4m^2}{1-z+k^2} \right)
\eaeq
and leads to the following maximal invariant gluon mass for given $z_1,z_2$:
\baeq\label{ksqmax}\eqalign{
 k^2\le k^2_m &= (1-z_1)(1-z_2) - \half\left[\, z_1z_2 - 4m^2
 - \sqrt{z_1-4m^2}\sqrt{z_2-4m^2}\, \right] \cr
&= \left\{ \frac{2\>(1-x+m^2)} {x-(2-x)y-4m^2 +\sqrt{[\,x^2-4m^2\,]
\>[\,(1\!-\!y)^2-4m^2\,]} } \right\} \cdot   (y_+ -y)(y-y_-)\,.
}\eaeq
As we shall see below, the structure of the matrix element is such that
essential
contributions emerge from the logarithmic region
$y_-\!\ll\! y \!\ll\! y_+$, as well as from the vicinity of the {\em lower}\/
limit, $(y\!-\!y_-)\sim y_-\!\sim\! m^2$.
Therefore for our purposes the following approximate expression suffices,
\baeq\label{ksqmaxapp}
k_m^2 = (1-x)(y-y_-) \>+\> \cO{m^2}\>,
\eaeq
which differs from the exact formula (\ref{ksqmax}) only in a tiny region
close to the {\em upper}\/ integration limit, $(y_+\!-\!y)\sim m^2$.

Two characteristic momentum scales emerge related to the limiting
values of the $y$-integration (\cf (\ref{twoscales}))
\baeq\label{twoscalesap}\eqal2{
\cQ^2(x) \>\>\equiv\>\>&        W^2\cdot (1-x) y_+
&  \>\>\approx\>\>  W^2x(1-x)\>;   \cr
\kappa^2(x)  \>\>\equiv\>\>&    W^2\cdot (1-x) y_-
&  \>\>\approx\>\>  M^2(1-x)^2/x   \>.
}\eaeq

\section{VASP Cross Sections}
The differential first order cross section integrated over angles of the final
$\QQ g$ system may be written as
\baeq\label{cs3}
\left\{\frac{d^2\sigma_{\QQ g}} {dz_1\,dz_2}\right\}_C \>=
\> \sigma^{\infty}  \cdot \frac{C_F\as}{2\pi}\> \Pi_C \>.
\eaeq
The subscript $C$ marks the production current (vector (V), axial vector (A),
scalar (S) or pseudoscalar (P)) and
$\sigma^{(\infty)}$
is the universal high-energy limit of
the Born transition probability,
$\mbox{Current}(C) \to \QQ$.

Straightforward calculation leads to the following expression for the
$\Pi$ factors\footnote{see \cite{DKS} for details}:
\bminiG{sig} \label{PiC}
 \Pi_C &=&   2\zeta_C \>\cS   \>\>+ \>\> \cH_C\>; \\
\label{cHV}
\cH_V &=& \k1k2 \>\>=\>\> (1-x)\cdot\frac1{y}\>+\> \frac{y}{1-x} \>, \\
\label{cHs}  \cH_A &=&    (1+2m^2)\cH_V + 4m^2 \>,\qquad
\cH_S  = \cH_P  = \cH_V +2 \>.
\emini
Here $\cS$ is the ``soft'' \br term
\baeq\label{cSdef}
\eqalign{
\cS \equiv  -\left(\frac{p_{1\mu}}{\kappa_1} -\frac{p_{2\mu}}{\kappa_2}
\right)^2
&\>=\>  \frac{z_1+z_2-1-2m^2}{(1-z_1)(1-z_2)}
-\frac{m^2}{(1-z_1)^2}-\frac{m^2}{(1-z_2)^2} \cr
&\>=\> \frac{x-2m^2}{1-x}\cdot \frac1y \>-\>
\frac{m^2}{y^2} \>-\> \frac{1-x+m^2}{(1-x)^2}  \,.
}\eaeq
Its contribution to each of the squared matrix elements is explicitly
proportional to corresponding $\zeta$-factor, the one that determines energy
dependence of the corresponding \QQ Born cross section,
\bminiG{Borns}
 && \sigma_\QQ = \sigma^{\infty} \cdot v\>\zeta_C(v)\>; \\
&& \label{zetaBorns}
\eqalign{
 \zeta_V &= \frac{3-v^2}2 = 1+2m^2\>, \cr
 \zeta_A &=  \zeta_S = v^2 = 1- 4m^2 \>, \qquad  \zeta_P = 1\>.
}\emini
As a result, the {\em normalized}\/ differential distribution can be written
in the following general form as a sum of a {\em universal}\/ ``$\cS$oft''
and a process dependent ``$\cH$ard'' contribution:
\baeq\label{general}
d^2w_C \equiv \left\{ \frac{d^2\sigma_{\QQ g}} {\sigma_{\QQ}}\right\}_C
= \frac{C_F\as}{2\pi\,v}\>  \left\{\> 2\cS
+   \zeta_C^{-1}\>\cH_{C} \>\right\}\> dx\,dy\,.
\eaeq

\paragraph{$z,\Theta_c$ Basis.}
Both structure of the matrix element and kinematics become particularly simple
in terms of the gluon energy fraction and the gluon angle in the
\QQ cms (see (\ref{udef})).  Making use of (\ref{usrel}) one gets
\bminiG{ztbas}
\label{cSinztbas}
\cS &=& \frac{4(1-z)}{z^2} \>  \frac{\beta^2-u^2}{(1-u^2)^2}
\>=\> \frac{4(1-z)}{z^2} \>
\frac{\beta^2\sin^2\Theta_c}{(1-\beta^2\cos^2\Theta_c)^2}
\frac{4m^2}{1-z+4m^2} \>; \\
\label{cHinztbas}
\cH_V &=& \frac4{1-u^2} -2 \>=\> \frac{4}{1-\beta^2\cos^2\Theta_c}-2 \>.
\emini
Invoking (\ref{zuphsp}), (\ref{general}) we obtain for the inclusive
gluon energy spectrum in the vector channel (see also \cite{BFK73})
\baeq\label{generalzu}
 {dw_V}   = \frac{C_F\as}{\pi}\, \frac{dz}{z} \frac{\beta}{v}
\int_{-1}^1 d\cos\Theta_c \left\{  {2(1\!-\!z)}  \>
 \frac{\beta^2\sin^2\Theta_c}{(1\!-\!\beta^2\cos^2\Theta_c)^2}
\> + \>   z^2\left[\,\frac{1}{1\!-\!\beta^2\cos^2\Theta_c}- \frac12 \,\right]
\zeta_V^{-1}  \right\} .
\eaeq
Our convention to call the two pieces of the matrix element ``soft'' and
``hard'' becomes clear now: in the soft gluon limit, $z\ll1$,
the second term is $z^2$ down compared to the first one.
In this representation the ``dead cone'' phenomenon is also manifesting:
the soft {\em classical}\/ term $\cS$ vanishes in the very forward directions
($\Theta_c, \pi\!-\Theta_c < \theta_0= m$).

In relativistic situation ($m\ll \half$) a collinear logarithmic enhancement
occurs and the two pieces of (\ref{generalzu})
participate in forming the GLAP splitting function,
$$
dw \> \propto\> \frac{d\Theta_c^2}{\Theta_c^2}
\> \left\{ {2(1-z)} \>+\> z^2 \right\}
 \frac{dz}{z}  \>\>\propto\>\> \frac{1+(1-z)^2}{z}\> dz \,.
$$
It is worthwhile to notice that $\cO{z^{-1}}$ and $\cO{1}$ parts of the
$q\to q+g(z)$ splitting function are process independent, while the $\cO{z}$
piece breaks factorization at the level of $\cO{m^2\ln m^2}$ correction.
Therefore the very notion of ``fragmentation function'' as a way of treating
the jet evolution independently of the production mechanism gets lost beyond
leading twist.

To stress logarithmic character of the angular integration one may
represent (\ref{generalzu}) as
\bminiG{generalzeta}
\label{wgeneralzeta}
 {dw_V}   = \frac{C_F\as}{\pi\,v}\> \frac{dz}{z}
 \int_{\eta_0}^1 \frac{d\eta}{\sqrt{1-\eta}}
 \left\{\,  {2(1-z)}  \> \frac{\eta-\eta_0}{\eta^2} \>\>+\>\>
 z^2\left[\,\frac{1}{\eta}- \frac12 \,\right] \zeta_V^{-1} \,\right\} \,,
\eeeq
where
\beeq
 \eta \equiv 1-u^2=1-\beta^2\cos^2\Theta_c \>\ge\>\eta_0\equiv
 \frac{4m^2}{1-z}\,.
\emini
For other production channels the second ``hard'' term in (\ref{generalzu}),
(\ref{wgeneralzeta}) should be changed according to (\ref{cHs}),
(\ref{zetaBorns}).

\section{Integration over Virtual Boson Mass}
Considering {\em inclusive}\/ characteristics
(\eg, such as the quark energy spectrum)
beyond the first order of perturbation theory one has to allow the gluon
to decay in the final state, that is to have a positive virtuality $k^2$,
and to integrate over the latter within the available phase space.
Actually, only such a combination of real and virtual gluon production
leads to a sensible physical answer.
First of all, an {\em exclusive}\/ real gluon production cross section is
clearly zero because of the standard  infinite (double logarithmic) Sudakov
form factor suppression.  Secondly, and more importantly,
a ``real'' gluon is an ill-defined object since its ``on-mass-shell''
interaction strength $\al_0$ may not be defined perturbatively
(which is a \underline{real} ``infrared catastrophe'' inherent for QCD).

Integration over $k^2$ leads to appearance of the running coupling in the
inclusive cross section.
This nice property discovered in the pioneering papers by Gribov and Lipatov
on partonic structure of logarithmic field theories\refup{GL}
is particularly helpful in the QCD context.
Here neither $\al_0$ in ``elastic'' radiation
nor small virtuality inelastic gluon decay systems are well defined.
Meanwhile, their combined action results in a legitimate $\as(k_m)$
at the Euclidean scale related to the maximal available gluon virtuality.

We shall demonstrate this correspondence taking care of subleading
effects which might be relevant within the adopted approximation.
To this end we write down a formal dispersion relation (with one subtraction)
for the running coupling
\bminiG{disrel}
\label{aldr}
 \al(Q^2) &\equiv& \al_0 \cdot \cZ_3(-Q^2) =
\al_0 + Q^2 \int_0^\infty \frac{dk^2}{k^2} \>\frac{\sigma(k^2)}{k^2 + Q^2} \>,
\eeeq
where $\sigma$ is related to discontinuity of
$\cZ_3$ at positive virtuality,
\beeq\label{spden}
\sigma(k^2) & \equiv&  \frac{\al_0}{2\pi\, i}
\left[\, \cZ_3(k^2-i\eps) - \cZ_3(k^2+i\eps)\, \right]
\>\equiv\> -\frac{\al_0}{2\pi}\> \mbox{Disc}\> \left\{\cZ_3(k^2)\right\}.
\emini
In QED $\cZ_3$ is nothing but the photon renormalization function
and $\al_0\approx 1/137$ --- the on-mass-shell coupling constant.

Strictly speaking, such an identification is true for an Abelian theory only.
In the QCD context the Ward identity between vertex and fermion propagator
corrections, $\cZ_\Gamma\cZ_q=1$, does not hold.
As well known, both the non-Abelian vertex renormalization correction
$\cZ^{(\mbox{\small na})}_\Gamma$ and the gluon propagator factor $\cZ_g$
participate (in a gauge dependent way) in forming the running $\as$.
So one has to look upon $\cZ_3$ in the dispersion relation (\ref{disrel})
as the gauge invariant product
$$
 \cZ_3 \>=\> \cZ^{(\mbox{\small na})}_\Gamma
             \cdot \cZ_g \cdot \cZ^{(\mbox{\small na})}_\Gamma\>.
$$
This subtlety does not affect the result, however.

Another motivation is to use the $n_f$ dependence as a gauge to pinpoint
the structure of the running $\as$ in the inclusive cross section\refup{BLM}.
Quark loops (in order $\al^2$) belong to $\cZ_g$ only, and employing
a ``quasi-Abelian'' relation (\ref{disrel}) becomes natural.

In the problem under consideration the following structure emerges,
\baeq\label{podint}
\eqalign{
& \al_0 M(z_1,z_2;0)
+  \int_0^{k_m^2}\frac{dk^2}{k^2}\> \sigma(k^2) \cdot  M(z_1,z_2;k^2) \cr
= & \left\{ \al_0 +  \int_0^{k_m^2}\frac{dk^2}{k^2}\> \sigma(k^2) \right\}
 M(z_1,z_2;0)
\>+\> \int_0^{k_m^2} dk^2\>  \sigma(k^2) \cdot \Delta M(z_1,z_2;k^2)\>,
}\eaeq
with $k_m^2=k_m^2(z_1,z_2)$  the maximal squared virtual boson mass
allowed for given $x, y$.
Here we have singled out the $k^2=0$ part of
the matrix element by writing
\baeq\label{Msplitgen}
 M(z_1,z_2;k^2) =  M(z_1,z_2;0) + k^2\cdot \Delta M(z_1,z_2;k^2) \> .
\eaeq
Now we may relate the characteristic integral in the first term in the rhs of
(\ref{podint})  with the dispersion formula (\ref{aldr}).
To this end we employ the following {\em exact}\/ representation of the
dispersion relation
\baeq
\al(Q^2)= \al_0 + Q^2 \int_0^\infty \frac{dk^2\>\sigma(k^2)} {k^2(k^2 + Q^2)}
= \al_0 + \int_0^{Q^2} \frac{dk^2}{k^2}\> \sigma(k^2) \>+\>
\int_0^1 \frac{dt}{1+t} \left[\, \sigma(Q^2/t) - \sigma(Q^2 t)\,\right] .
\eaeq
Perturbatively, $\sigma$ is of the order of $\al^2$.
As a result, the last finite integral term constitutes a
{\em next-to-next-to-leading}\/ correction to the main one.
To see this one has to view $\sigma$ as a slowly varying (logarithmic) function
of its argument\footnote{which is true everywhere except in the very vicinity
of a fermion threshold} and to perform the Taylor expansion in $\ln Q^2$
to obtain
$$
 \int_0^1 \frac{dt}{1+t} \left[\, \sigma(Q^2/t) - \sigma(Q^2 t)\,\right]
 \>=\> -2 \sigma'(Q^2) \int_0^1\frac{dt}{1+t} \ln t  + \cO{\sigma'''}
 \>\approx\> \frac{\pi^2}{6}\,  \sigma'(Q^2) \>.
$$
This exercise demonstrates a close correspondence between the spectral density
(\ref{spden}) and the $\beta$-function\refup{SCHW}:
\bmini
 \al(Q^2) &=&  \al_0 + \int_0^{Q^2} \frac{dk^2}{k^2}\> \sigma(k^2)
 \>+\>  \frac{\pi^2}{6}\,  \sigma'(Q^2) + \ldots \>; \\
 \sigma(Q^2) &=&  \al'(Q^2) -  \frac{\pi^2}{6}\, \al'''(Q^2) + \cO{\al^{(V)}}
\,.
\emini
Thus we express perturbatively the structure in curly brackets of
(\ref{podint}) in terms of the running coupling at the Euclidean point
$Q^2=k_m^2$:
\baeq\label{altil}
\tilde{\al}(k_m^2) \equiv
\left\{ \al_0 +  \int_0^{k_m^2}\frac{dk^2}{k^2}\> \sigma(k^2) \right\}
\>=\> \al(k_m^2) - \frac{\pi^2}{6}\, \al''(k_m^2) + \ldots
\eaeq

\section{Running Coupling in the First Order Spectrum}
It is straightforward to calculate the exact $\QQ g$ matrix  element
with account of a non-zero virtual gluon mass $k^2>0.$
(\ref{general}) gets modified as follows:
\bminiG{ksqme}
2\cS   &\Longrightarrow& 2\cS^{(0)} \>-\>
 k^2 \left( \frac{1}{(1-z_1)^2} + \frac{1}{(1-z_2)^2} \right)   ; \\
\cH_V  &\Longrightarrow& \cH_V^{(0)} \>+\> 2k^2
\frac{ z_1+z_2+k^2 }{(1-z_1)(1-z_2)} \>,
\emini
where $\cS^{(0)}$ and $\cH_V^{(0)}$ are given by the original
``on-mass-shell-boson'' expressions (\ref{cSdef}) and (\ref{cHV}) respectively.
In terms of (\ref{Msplitgen}),
\bminiG{Msplit} \label{Msp1}
\Delta\left\{\, 2\cS \,\right\} &=& -\frac{1}{y^2}\> - \frac{1}{(1-x)^2}\>;\\
\label{Msp2}
\Delta\left\{\, \cH_V \,\right\} &=&  \frac{2(1+x-y+k^2)}{(1-x)\,y} \>.
\emini

\subsection{$k^2=0$ part of the matrix element.}
To evaluate the radiator (\ref{radiator}) we start by considering
the first term in (\ref{podint}) proportional to $M(z_1,z_2;0)$.
One has to perform the $y$-integral with the factor $\tilde{\al}$
that emerges after integration over virtual boson mass, see (\ref{altil}),
\baeq
 \int_{y_-}^{y_+} dy \>\tilde{\al}(k_m^2(x,y)) \cdot
\left\{\> 2\cS^{(0)} + \zeta_V^{-1}\cH^{(0)} \>\right\} ,
\eaeq
with $k_m^2$ given by the approximate expression (\ref{ksqmaxapp}). Invoking
(\ref{cSdef}), (\ref{cHV}) we split the integrand into three pieces as follows:
\bminiG{cSHspl} \label{cSHlog}
\left\{ 2\cS^{(0)} + \zeta_V^{-1}\cH^{(0)} \right\} &=&
 \left\{\> 2\frac{x-2m^2}{1-x} + \zeta_V^{-1}(1-x)\> \right\} \frac1y \\
\label{cSHsin} &-& 2\frac{m^2}{y^2}  \\
\label{cSHfin} &-& 2\frac{1-x+m^2}{(1-x)^2} \>+\> \zeta_V^{-1}\frac{y}{1-x} \,.
\emini

\paragraph{Logarithmic Piece (\ref{cSHlog}).}
This is the leading contribution to the radiator, in which the $y$-integration
is logarithmic (collinear log):
\baeq\label{logex}
  \int_{y_-}^{y_+} \frac{dy}{y}\> \tilde{\al}(y-y_-)\>.
\eaeq
Here we have explicitly shown the essential $y$-dependence of the coupling
factor  $\tilde{\al}$.  The chain of approximations follows:
\baeq\eqalign{
 (\ref{logex}) &=  \int_{y_-}^{y_+} \frac{dy}{y}\>
\left( \tilde{\al}(y) + {\tilde{\al}}'(y)\ln\frac{y-y_-}{y} + \ldots \right)
\approx \int_{y_-}^{y_+} \frac{dy}{y}\> \tilde{\al}(y)
 +  {\tilde{\al}}'(y_)  \int_{\frac{y_-}{y_+}}^1 \frac{ds}{s}\ln(1-s) \cr
& \approx \int_{y_-}^{y_+} \frac{dy}{y}\> \tilde{\al}(y)
 -\frac{\pi^2}{6}\> {\tilde{\al}}'(y_-)
\>\> = \>\>  \int_{\kappa^2}^{\cQ^2} \frac{dt}{t}\> {\al}(t)
 -\frac{\pi^2}{6}\> {\al}'(\cQ^2) \>+\> \cO{\al^3(\kappa^2) + \al^3(\cQ^2)},
}\eaeq
where (\ref{altil}) has been used.
We notice that a potential second order contribution  $\al'\propto \al^2$
at the {\em low}\/ scale $\kappa^2\sim M^2$ cancels out, and the remaining
term ${\al}'(\cQ^2)\propto \al^2(W^2)$ is comparable with the two-loop
correction to the {\em hard cross section}\/ (coefficient function)
and must be neglected within our approximation.

Finally,
\baeq\label{p1}
\int dy \>(\ref{cSHlog}) =\left(\> 2\frac{x-2m^2}{1-x} +
\zeta_V^{-1}(1-x)\>\right)
\int_{\kappa^2}^{\cQ^2} \frac{dt}{t}\> \al(t)\>.
\eaeq

\paragraph{Singular Piece (\ref{cSHsin}).}
The second term of (\ref{cSHspl}) originates from the dead cone subtraction.
It is explicitly proportional to $m^2$ but this suppression gets compensated
by the singular behaviour in $y$. Corresponding $y$-integral is concentrated
in a tiny region $(y-y_-)\sim y_-\propto m^2$:
\baeq
 \int_{y_-}^{y_+} \frac{dy}{y^2}\> \tilde{\al}(y-y_-)
= \al(y_-)\int_{y_-}^{y_+} \frac{dy}{y^2}
+ \al'(y_-)\int_{y_-}^{y_+}\frac{dy}{y^2}\ln\frac{y-y_-}{y_-} +\cO{\al''(y_-)}
\eaeq
The $\cO{\al^2(y_-)}$ term vanishes in the relativistic approximation:
$$
\int_{y_-}^{y_+} \frac{dy}{y^2} \ln\frac{y-y_-}{y_-}
= \int_0^{\frac{y_+-y_-}{y_-}} \frac{ds}{(1+s)^2} \ln s
\approx \int_0^{\infty} \frac{ds}{(1+s)^2} \ln s  =0\,.
$$
Thus, similarly to what has happened to the logarithmic term (\ref{cSHlog})
discussed above, a potential $\cO{\al^2(\kappa)}$ contribution is absent.
One is left with a pure $\al(\kappa^2)$ correction to the radiator:
\baeq\label{p2}
\int dy \>(\ref{cSHsin}) =  -2m^2\left\{ \frac{1}{y_-}-\frac{1}{y_+}\right\}
\al(\kappa^2)
= - 2\frac{\sqrt{x^2-4m^2}}{1-x}\> \al(\kappa^2)\,.
\eaeq

\paragraph{Finite Piece (\ref{cSHfin}).}
Corresponding $y$-integral is collinear safe (\ie\ finite in the $m^2\!=\!0$
limit) and constitutes $\al(W^2)$ correction to the hard cross section.
Here one extracts the coupling at the upper limit, $y\sim y_+$,
and the $y$-integration becomes trivial (see (\ref{usrel2})):
\baeq\label{p3}\eqalign{
\int dy \>(\ref{cSHfin}) &= \al(\cQ^2) \left\{
\>- 2\frac{(1-x+m^2)(y_+-y_-)}{(1-x)^2}
\>+\> \zeta_V^{-1}\frac{\half(y_+^2- y_-^2)}{1-x}
 \right\} \left[\>1+\cO{\al(\cQ^2)} \>\right] \cr
&\approx\> \al(\cQ^2) \sqrt{x^2-4m^2} \left\{  \>- \frac{2}{1-x}
\>+\> \zeta_V^{-1}\frac{x-2m^2}{2(1-x)}
\left(\frac{1-x}{1-x+m^2}\right)^2\right\} \,.
}\eaeq

\subsection{$\Delta$ part of the matrix element.}
{}From the first sight, corrections (\ref{Msplit}) to the $\QQ g$ matrix
element
proportional to virtual boson mass look negligible: corresponding $k^2$
integration is no longer logarithmic, $k^2\sim k_m^2$, and the result is
of the order of $\sigma(k_m^2)\propto \al^2(k_m^2)$.

This expectation indeed comes true for the ``hard'' correction term
(\ref{Msp2}) as well as for the second piece of (\ref{Msp1}).
However, the first contribution to the ``soft'' correction term (\ref{Msp1})
is {\em over-singular}\/ in $y$, as a result of which singularity
logarithmic behaviour gets restored and a contribution to the
two-loop anomalous dimension emerges.

Indeed,
\bminiG{sigtil}
&& \eqalign{ \int_0^{k_m^2} dk^2 \sigma(k^2) & = k_m^2 \left\{ \sigma(k_m^2)
+  \sigma'(k_m^2) \int_0^{k_m^2} \frac{dk^2}{k_m^2} \ln \frac{k^2}{k_m^2}
+ \ldots \right\} \>\> \equiv\>\>  k_m^2\cdot  \tilde{\sigma}( k_m^2 )\,,
}\eeeq
where
\beeq
  \tilde{\sigma}  \>=\>  \al'  -  \al''  \>+\> \cO{\al'''} \,.
\emini
Invoking (\ref{ksqmaxapp}) for $k_m^2$ one arrives at the following
expression for the $\Delta$-correction:
\baeq
-(1-x) \int_{y_-}^{y_+} \frac{dy}{y^2} (y-y_-)  \cdot \tilde{\sigma}(y-y_-)
\eaeq
An approximate evaluation follows:
\baeq
\eqalign{
 \int_{y_-}^{y_+} & \frac{dy}{y^2} \,(y-y_-) \tilde{\sigma}(y-y_-)
=   \int_{y_-}^{y_+} \frac{dy}{y^2} \,(y-y_-)  \tilde{\sigma}(y)
\>\>+\>\>  \cO{\tilde{\sigma}'(y_-)}   \cr
&\approx \int_{y_-}^{y_+} \frac{dy}{y} \, \tilde{\sigma}(y)
-  \tilde{\sigma}(y_-)
= \left[\,\al(\cQ^2)- \al(\kappa^2) - \al'(\cQ^2)+ \al'(\kappa^2) \,\right]
\>-\> \left[\, \al'(\kappa^2) + \ldots \right].
}\eaeq
Again, as before, the $\al^2(\kappa^2)$ contribution cancels,
and one finally gets the correction
\baeq\label{Delcor}
 \int dy\, \int dk^2 \sigma(k^2)\> \Delta M(x,y; k^2)
 \>=\> -(1-x) \left[\,\al(\cQ^2)- \al(\kappa^2) \,\right] \>\> + \>\>
 \cO{\al^2(\cQ^2) + \al^3(\kappa^2)}\>.
\eaeq
Combining (\ref{p1}), (\ref{p2}), (\ref{p3}) and (\ref{Delcor}) we finally
arrive at the expression (\ref{radiator}) for the perturbative radiator.
\eAPP

\newpage
\bAPP{B}{Second loop Anomalous Dimension $\Delta^{(2)}$}

\section{AD (dimensional regularization) }
The two-loop anomalous dimension (AD) has been derived in the framework of
dimensional regularization approach in \citm{CFP}{secloop}.
In notations of Curci, Furmanski and Petronzio\refup{CFP} the non-singlet AD
corresponding to $Q\to Q$ transition in {\em space-like}\/ evolution reads
(eqs.~(4.50) -- (4.54) of \cite{CFP})
\baeq\label{sl}
\eqalign{
C_F^{-1}\hat{P}_{qq}(x,a) &=  aP(x) +  a^2\ad^{(2)}(x) \>, \qquad
P(x) \equiv \frac{1+x^2}{1-x} \>; \cr
\ad^{(2)} &=   C_F\,P_F(x) + C_A\, P_G(x) + n_f T_R\, P_{n_f}(x)\>,
}\eaeq
where
\bminiG{slall}
\label{sl1}
P_F &=& -2P(x)\ln x\ln (1-x) -\left(\frac3{1-x} + 2x\right)\ln x
-\half(1+x)\ln^2x - 5(1-x) \\
\label{sl2}
P_G &=& P(x)\left[\, \half\ln^2x + \frac{11}{6}\ln x +\frac{67}{18}
-\frac{\pi^2}{6} \,\right] + (1+x)\ln x + \frac{20}{3}(1-x) \\
\label{sl3}
P_{n_f} &=& -\frac23\left[\, P(x)\left(  \ln x + \frac53 \right) +
2(1-x)\,\right] .
\emini
Let us mention an extra contribution due to $Q\to \bar{Q}+QQ$ transition
which reads
\bminiG{extrsing}
C_F^{-1}\hat{P}_{q\bar{q}}(x,a) =  a^2 \left(C_F - \half C_A \right)
\left\{\> 2P(-x)S_2(x) +2(1+x)\ln x +4(1-x) \>\right\},
\eeeq
with
\beeq
 S_2(x) \>\equiv\> \int_{x(1+x)^{-1}}^{ (1+x)^{-1}} \>
\frac{dz}z \> \ln\frac{1-z}{z} \>.
\emini
This contribution (which formally belongs to the {\em non-singlet}\/ AD, see,
\eg, \cite{Guido}) vanishes as $(1\!-\!x)^2$ at large $x$, is colour
suppressed and numerically negligible.  In this paper it has been disregarded
together with the {\em singlet}\/ (sea) contribution to heavy quark yield.

An algebraic massage leads to the following representation for the
quark evolution kernel\footnote{The ``+'' prescription is implicit.}:
\baeq\label{adshort}
 C_F^{-1}\hat{P}_{qq}(x,a) =  \left[\,a+\cK a^2\,\right] P(x)
 \>+\>  a^2 \left\{\sigma \cdot C_F\cV(x) + \cR(x)  \right\}
 \>-\> a' \left[\, P(x)\ln x + 2(1-x)\, \right].
\eaeq
The first term here collects the one-loop AD with the part of
$\ad^{(2)}$ correction explicitly proportional to $P(x)$ with the number
$\cK$ given by (\ref{cKdef}) above. These two combine forming a ``physical''
coupling in terms of the $\MSbar$ one according to (\ref{phco}).
This term totally absorbs the  $(1-x)^{-1}$ singularity of the evolution
kernel: $\cV(x)\propto \ln(1\!-\!x)$ and $\cR(x)$ vanishes as
$(1\!-\!x)$ in the quasi-elastic limit.

\noindent
(The very last term in (\ref{adshort}) proportional to the derivative of the
running coupling,
\baeq
 a' \>\equiv\> \frac{d}{d\ln W^2}\, \frac{\as(Q)}{2\pi}
 \>=\> - \left( \rat{11}6 N_c - \rat23 T_Rn_f\right)\, a^2 \>+ \ldots\,,
\eaeq
does not count, as it is an artefact of the dimensional
regularization scheme; see below.)

The ``true'' second loop AD in curly brackets of (\ref{adshort})
consists of two contributions.
The first one is
\bminiG{GLV} \eqalign{
  \cV(x) &= -\left[\> P(x)\left(\ln\frac{x}{(1-x)^2} -\rat{3}{2}\right)
   + (1-x) -\half(1+x)\ln x \>\right]\ln x \cr
&=  \int_0^1 dz \int_0^1 dy \,\delta(x-yz) \> \left\{P(y)\right\}_{+}\,
P(z)\ln z \>.
}\eeeq
So defined, $\cV$ has a very simple form in the moment representation, namely,
\beeq\label{GLVj}
\cV(j) \equiv \int_0^1 dx \, x^{j-1} \>  \cV(x)
=    P_j\, \frac{d}{dj} P_j\>. \qquad
\left(\> P_j\equiv  \int_0^1 dx \,P(x)_+ = \int_0^1 dx
\left[\, x^{j-1} -1\,\right]
\frac{1+x^2}{1-x}\>\right)
\emini
(Obviously, $[\cV(x)]_+=\cV(x)$.)
This (and only this) contribution changes (acquires an opposite sign)
when time-like evolution is considered\refup{CFP}.
Therefore (\ref{adshort}) holds for both channels with
$\sigma =\pm 1$ referring to time- and space-like evolution correspondingly.
The origin of the $\cV$ term in the two-loop kernel may be traced backed to
a simple kinematical difference between annihilation and scattering
channels\refup{CFP}.
It has been argued\refup{DESY} that a mismatch between the two-loop
$\ee$ and DIS anomalous dimensions would disappear
(and thus the Gribov-Lipatov relation\refup{GL} would be rescued)
if one considered scaling violation of ``pseudo-moments''
in $x$ evaluated for a fixed value of
$\left\{xW^2\right\}_{\mbox{annih.}}=\left\{Q^2/x\right\}_{\mbox{scatt.}}$

Another term of the second loop AD is
\baeq
  \cR(x) =  (\half C_A -C_F)  P(x)\ln^2 x
 - 5C_F\left[\,  \half(1\!+\!x) \ln x + (1\!-\!x)\, \right]
 + C_A \left[\, (1\!+\!x)\ln x + {3}(1\!-\!x) \, \right] .
\eaeq
Similarly to the one-loop splitting function $P(x)$,
it obeys the reciprocity relation\refup{GL}
\bmini
 -x\cR(x^{-1}) =  \cR(x)\,,
\eeeq
and, as a result, stays the same for time- and space-like evolution.
Contrary to this,
\beeq
 -x\cV(x^{-1}) = [\,-1\,]\cdot \cV(x)\,.
\emini

\section{CF (dimensional regularization) }
$\cO{a}$ correction to the hard cross section\refupd{CFP}{AEMP}
(coefficient function; CF)
has to be taken into consideration together with the two-loop AD since
neither CF nor AD is a scheme independent quantity beyond leading logs.

It reads (see eqs.~(7.4),~(7.10) of \cite{CFP})
\bminiG{CFmsbar}  \label{CF1}
C_{2}^A  &=& \delta(1-x) \>+\> aC_F \left\{ \left[\>\frac{1+x^2}{1-x}
\left(\ln\frac{1-x}{x} -\rat34 \right)
 + \rat14(9+5x) \>\right]_{+} \right.\nonumber\\
&+& \left.\frac{3(1+x)^2}{1-x}\ln x -\rat72(1+x)+\pi^2\delta(1-x)\>\right\};\\
\label{CF2}     C_{L}^A  &=& aC_F\>.
\emini
The first line of (\ref{CF1}) is the CF for the scattering
channel ($C_2^S$); the second line accounts for their difference.
After simple manipulations one arrives at
\baeq\label{CF12}
 C_{2}^A \>=\> (1-\rat32 aC_F)\delta(1\!-\!x) + aC_F\left[\,
 P(x)\left(\ln[x^2(1-x)] -\rat34\right) -\rat14 (5+9x)\,\right]_{+} \>.
\eaeq
In the cross section integrated over total $\QQ g$ production angle
(see (\ref{angint}))  the combination $C_2^A + 3C_L^A$ emerges.
With account of the longitudinal contribution the $\delta(1\!-\!x)$-term
gets modified,
\baeq\eqalign{
C^A(x,a) \equiv C_2^A + 3C_L^A &\>=\> (1-\rat32 aC_F)\delta(1\!-\!x)  +
aC_F [\,(\ref{CF12})\,]_+  \>+\>  3\,aC_F \cr
&\>=\> (1+\rat32 aC_F)\delta(1\!-\!x) + aC_F [\,(\ref{CF12})+3\,]_+\>.
}\eaeq
Constructing the CF that describes quark distribution
{\em normalized}\/ by the total cross section,
\baeq
 C_t(x,a) =  (1+\rat32 aC_F)^{-1} \cdot C^A(x,a) \>,
\eaeq
one gets, with $\cO{a}$ accuracy, the answer which may be presented
in the following form:
\baeq\label{Cfin}\eqalign{
C_t(x,a) = \delta(1\!-\!x) &\>\>+\>\>
aC_F\left[\,P(x)\left(\ln[x(1-x)] -\rat34\right) -\rat14(1+x)\,\right]_{+} \cr
&\>\>+\>\> aC_F\left\{\> P(x)\ln x + 2(1-x)\> \right\}_{+}  \>.
}\eaeq

\section{Scaling Violation Rate}
Scaling violation rate is an observable determined by the
scheme invariant combination
\baeq
D' \equiv \frac{d}{d\ln W^2}D(x,W^2) \>\Longrightarrow\>
\hat{P_{qq}}(x,a(W^2)) \>+\> \frac{d}{d\ln W^2}\, C_t(x,a(W^2))\,.
\eaeq
Combining (\ref{adshort}) and (\ref{Cfin}) one arrives at
\footnote{The ``+'' prescription is implicit.}
\baeq\label{scviol}
\eqalign{
D' \>\Longrightarrow\> \left[\,a_{\MSbar}+\cK a^2\,\right] C_FP(x)
&\>+\>  a^2C_F \left\{\>C_F\cV(x) + \cR(x)\>  \right\} \cr
& \>+\>a' C_F\left[\,P(x)\left(\ln[x(1-x)] -\rat34\right)-\rat14(1+x)\,\right],
}\eaeq
where the $\MSbar$ origin of the running coupling has been stressed
in the relevant first order term.

This result has to be compared with the scaling violation rate as described
by (\ref{radiator}).  To this end one evaluates the $\ln W^2$ derivative
of the relativistic radiator (\ref{radrel}) to obtain
\baeq\label{Dpr1}
D' \>\Longrightarrow\>  \left\{ a(\cQ^2)C_FP(x) -a'C_F(1-x)
+ a^2C_F\Delta^{(2)} \right\}
\>+\> a'C_F \left\{ \frac{-2x}{1-x} + \frac{x^2}{2(1-x)} \right\} .
\eaeq
In the leading term one has to expand the coupling of a composite argument
$\cQ^2$ near $W^2$ as follows,
\baeq
 a(\cQ^2) = a(W^2) + a' \ln[x(1-x)] + \ldots
\eaeq
Adding together terms proportional to $a'$
one observes correspondence with a relevant part of (\ref{scviol}):
$$\eqalign{
  -\frac{2x}{1-x} = -P(x) + (1\!-\!x) \>, \qquad &
  \frac{x^2}{2(1\!-\!x)} = \rat14 P(x) -\rat14(1+x) \>; \cr
  \ln[x(1\!-\!x)] -(1\!-\!x) -\frac{2x}{1-x} + \frac{x^2}{2(1\!-\!x)}
\>&= \> P(x)\left(\ln[x(1\!-\!x)] -\rat34\right) -\rat14(1+x) \,.
}$$
Identifying
\baeq
(\ref{scviol})  \>=\> (\ref{Dpr1})
\eaeq
we arrive at the final result which relates an ``effective''
(dispersion scheme motivated) coupling $\aef$  with  $a_{\MSbar}$
and gives an expression for the ``true'' second loop
correction to the radiator $\tilde{\Delta}^{(2)}$ we were aiming at:
\baeq
\aef C_FP(x) \>+\> a^2C_F\tilde{\Delta}^{(2)}
\>\>=\>\> \left[\,a_{\MSbar}+\cK a^2\,\right] C_FP(x)
\>+\>  a^2C_F \left\{\,C_F\cV(x) + \cR(x)\,  \right\} .
\eaeq
\eAPP

\newpage
\def\labelenumi{[\arabic{enumi}]}

\noindent
{\Large\bf References}
\ben
\item\label{frs}
for reviews see, \eg, \\
P. M\"attig, Talk at the $4^{th}$ Int. Symp. on
Heavy Flavour Physics; Bonn preprint,\\
Bonn-HE--91--19 (1991); \\
T. Behnke in: Proc. of the 26th Int. Conf.
on High Energy Physics, Dallas, Texas, 1992,
ed. J.R Sanford, AIP Conference Proceedings, vol. I, p.859; \\
see also T. Behnke to be published in: Proc. of the Conf. QCD-94,
Montpellier, France, 1994; \\
P.S. Wells, preprint CERN-PPE/94--203 (1994).

\item\label{exp}
for recent publications see, \eg, \\
OPAL Collaboration, R. Akers \others, \zp{60}{199}{93} ;\\
SLD Collaboration, K. Abe \others, \prl{72}{3145}{94} ; \\
OPAL  Collaboration, R. Akers \others, \zp{61}{209}{94} ; \\
A.De Angelis \others, DELPHI Report, DELPHI 94--94, PHYS 411 ; \\
ALEPH Collaboration, D. Buskulic \others, \zp{62}{1}{94},
\ib{C62}{179}{94}.

\item\label{r3}
Yu.L. Dokshitzer, V.S. Fadin and V.A. Khoze, \pl{115}{242}{82}; \\
\zp{15}{325}{82}.

\item\label{r4}
Yu.L. Dokshitzer, V.A. Khoze and  S.I. Troyan, in:
Proc. of the 6th Int. Conf. on Physics in Collisions,
p.417, ed. M. Derrick, World Scientific, Singapore, 1987.

\item\label{r5}
Yu.L. Dokshitzer, Light and Heavy Quark Jets in Perturbative QCD,
in: {\em Physics up to 200 \TeV}\/,
Proc. of the Int. School of Subnuclear Physics,
Erice, 1990, vol.28, Plenum Press, New York, 1991.

\item\label{r7}
Yu.L. Dokshitzer, V.A. Khoze,  S.I.Troyan,
{\em J.Phys.G: Nucl.Part.Phys.}\/ {\bf 17}(1991)  1481, 1602.

\item\label{DKThq}
Yu.L. Dokshitzer, V.A. Khoze,  S.I.Troyan,
``Specific Features of Heavy Quark Production. \\
 I.~Leading quarks'', Lund preprint LU~TP~92--10;\\
 Yu.L. Dokshitzer,
in:  {\em Perturbative QCD and Hadronic Interactions}\/,
Proc. of the 27th Recontres de Moriond, p.259, ed. J.Tran Thanh Van,
Editions Fronti{\`e}res,  Gif-sur-Yvette, 1992.

\item\label{r8}
V.A. Khoze, in: Proc. of the 26th Int. Conf. on High Energy Physics,
Dallas, Texas, 1992, ed. J.R. Sanford,
AIP Conference Proceedings, vol.~II, p.~1578;
Durham preprint DTP/93/78.

\item\label{fragm}
C.~Peterson \others, \pr{27}{105}{83};  for review see, \eg, \\
M.~Bosman \others, {H}eavy {F}lavours, in:
Proc. of the Workshop on $Z$ physics at LEP,
CERN Report 89--08, ed.~G. Altarelli, R. Kleiss and C. Verzegnassi,
volume~1, p.267, 1989; \\
J.H. K\"uhn and P.M. Zerwas, in Advanced Series in Directions in
High Energy Physics, ``Heavy Flavours'', ed.
A.J.~Buras and M.~Lindner, World Scientific, Singapore 1992, p.~434.

\item\label{MeNa}
B. Mele and P. Nason, \pl{245}{635}{90};  \np{361}{626}{91}.

\item\label{CoNa}
G. Colangelo and P. Nason, \pl{285}{167}{92}.

\item\label{MLLA_LPHD}
Yu.L. Dokshitzer, V.A. Khoze, A.H. Mueller and S.I. Troyan,
Basics of Perturbative QCD, ed. J.Tran Thanh Van,
Editions Fronti{\`e}res,  Gif-sur-Yvette, 1991.

\item\label{BK}
V.N. Baier and V.A. Khoze,  \spj{21}{629}{65};
preprint NSU, Novosibirsk, 1964.

\item\label{thresh}
V.S. Fadin and V.A. Khoze, \jl{46}{525}{87}; \sjnp{48}{309}{88}.

\item\label{Guido}
G. Altarelli, \prep{81}{1}{82}.

\item\label{DS}
Yu.L.Dokshitzer, D.V. Shirkov,
``On Exact Account of Heavy Quark Thresholds in Hard Processes'',
Lund preprint LU~TP~93--19; unpublished.

\item\label{Luca}
L. Trentadue, in:  {\em Perturbative QCD and Hadronic Interactions}\/,
Proc. of the 27th Recontres de Moriond, p.167,
ed. J.Tran Thanh Van, Editions Fronti{\`e}res,  Gif-sur-Yvette, 1992.

\item\label{GL}
 V.N. Gribov and L.N. Lipatov, \sjnp{15}{438, 675}{72}.

\item\label{CFP}
G. Curci, W. Furmanski and R. Petronzio, \np{175}{27}{80}.

\item\label{AEMP}
G. Altarelli, R.K. Ellis, G. Martinelli and S.Y. Pi,
\np{160}{301}{79}.

\item\label{secloop}
W. Furmanski and R. Petronzio, \pl{97}{437}{80}, \zp{11}{293}{82}; \\
J. Kalinowski, K. Konishi, P.N. Scharbach and T.R. Taylor, \np{181}{253}{81};\\
E.G. Floratos, C. Kounnas and R. Lacaze, \np{192}{417}{81}.

\item\label{CMW90}
S. Catani, G. Marchesini and B.R. Webber,  \np{349}{635}{91}.

\item\label{BLM}
S.J.~Brodsky, G.P. Lepage and P.R.~MacKenzie,
\pr{28}{228}{83}; \\
for a recent review see
S.J.~Brodsky and H.J. Lu, SLAC-PUB-6683 (1994).

\item\label{LBK}
F.E. Low, \pr{110}{974}{58} ; \\
T.H. Burnett and N.M. Kroll, \prl{20}{86}{68}.

\item\label{DKS}
Yu.L. Dokshitzer, V.A. Khoze and W.J. Stirling, \np{428}{3}{94}.

\item\label{BFK73}
V.N. Baier, V.S. Fadin and V.A. Khoze,
\np{65}{381}{73}.

\item\label{leading}
Ya.I.Azimov, L.L.Frankfurt,  V.A.Khoze, Preprint LNPI--222, 1976;
\quad in: {\em Proc. of the XVIII Int. Conf. on High Energy Physics, Tbilisi,
1976}, Dubna, II:B10, 1977;\\
J.Bjorken,  \pr{17}{171}{78};\\
M.Suzuki, \pl{71}{139}{77}.

\item\label{Jaffe}
R.L. Jaffe and L. Randall, \np{412}{79}{94}.

\item\label{BH}
 V.N. Gribov, Possible Solution of the Problem of Quark Confinement,
 LU~TP~91--7, 1991.

\item\label{banda}
I.I.~Bigi, M.A.~Shifman, N.G.~Uraltsev and A.I.~Vainshtein,
\pr{50}{2234}{94}; \\
M.~Beneke and V.M.~Braun, \np{426}{301}{94}.

\item\label{Voloshin}
 M.B.~Voloshin and Yu.M.~Zaitsev,
\spu{30}{553}{87}.

\item\label{preprint}
Yu.L. Dokshitzer, V.A. Khoze,  S.I.Troyan,
``Specific Features of Heavy Quark Production. \\
 II.~LPHD Approach to Heavy Particle Spectra'', Lund preprint LU~TP~94--23.

\item\label{MikeS}
M.H. Seymour, \zp{63}{99}{94}.

\item\label{PeterCC}
OPAL Collaboration, R.Akers et al.,
CERN-PPE/94--217,
CERN-PPE/95--058.

\item\label{shapes}
Yu.L. Dokshitzer and B.R. Webber, \pl{352}{451}{95}.

\item\label{CTW}
       J. Kodaira and L. Trentadue, \pl{112}{66}{82}; \\
       C.T.H. Davies, J. Stirling and B.R. Webber, \np{256}{413}{85}; \\
       J.C. Collins, D.E. Soper and G. Sterman, \np{250}{199}{85}; \\
       S. Catani, E. d'Emilio and L. Trentadue, \pl{211}{335}{88}.
\item\label{CTS}
       G. Sterman, \np{281}{310}{87};\\
       S. Catani and L. Trentadue, \pl{217}{539}{89}; \np{327}{323}{89}; \\
       \np{353}{183}{91}.

\item\label{KT}
       J. Kodaira and L. Trentadue, \pl{123}{335}{82};
       preprint SLAC-PUB-2934 (1982).

\item\label{CTWT}
       S. Catani, G. Turnock, B.R. Webber and L. Trentadue,  \pl{263}{491}{91}.

\item\label{hjm}
       S. Catani, G. Turnock and  B.R. Webber, \pl{272}{368}{91}.

\item\label{pion}
       Yu.L. Dokshitzer, D.I. Dyakonov and S.I. Troyan,
       \pl{84}{234}{79}; \\ \prep{C58}{270}{80};\\
       A. Bassetto, M. Ciafaloni and G. Marchesini, \np{163}{477}{80}; \\
       G. Curci and M. Greco, \pl{92}{175}{80}.

\item\label{PP}
 G.Parisi and R.Petronzio, \np{154}{427}{79};\\
 B.R. Webber and P. Rakow, \np{187}{254}{81}.

 \item\label{pheno}
 Yu.L. Dokshitzer, V.A. Khoze and S.I. Troyan,  \zp{55}{107}{92}.

\item\label{CKL}
 J. Chyla, A. Kataev, S.A. Larin, \pl{267}{269}{91}.

\item\label{MStev}
A.C. Mattingly and P.M. Stevenson, \prl{69}{1320}{92}.

\item\label{PMS}
P.M. Stevenson, \pr{23}{2916}{81}.

\item\label{SCHW}
J. Schwinger,  {\em Proc.Nat.Acad.Sci.USA}\/ \underline{71} (1974) 5047.

\item\label{DESY}
Yu.L. Dokshitzer, Talk given at the HERA Workshop, DESY,
September 1993, unpublished.

\een
\end{document}